\documentclass[twocolumn,showpacs,amsmath,prd,amssymb,showkeys,superscriptaddress,aps,longbibliography,nofootinbib]{revtex4-2}

\pdfoutput=1

\usepackage{graphicx}
\usepackage{dcolumn}
\usepackage{mathtools}
\usepackage{scalerel}
\usepackage{bm}
\usepackage{mathrsfs}
\usepackage{amstext}
\usepackage{amsmath}
\usepackage{amssymb}
\usepackage{textcomp}
\usepackage{multirow}
\usepackage{lineno}

\usepackage[unicode=true,pdfusetitle,
 bookmarks=true,bookmarksnumbered=false,bookmarksopen=false,
 breaklinks=false,pdfborder={0 0 1},backref=false,colorlinks=true, citecolor=blue]
{hyperref}

\def\nuebar{\bar{\nu}_e} 

\def\nuAel{\nu {A}_{el}}
\def\munu{\mu_{\nu}}
\def\sigmanuA{\sigma_{\nuAel}}
\def\muB{\mu_{B}}

\def\DARpi{\pi^{\rm DAR}}
\def\q2{q^2}
\def\Enu{{\rm E}_{\nu}}
\def\MN{\rm M }
\def\TNR{\rm T_{\rm nr}}
\def\T0{\rm T_{thr}}  

\def\xe135{^{135}{\rm Xe}}

\def\CRV{{\rm CR}^{\mbox{-}}}
\def\ACV{{\rm AC}^{\mbox{-}}}

\def\BigDet{{\rm D_{70}}}
\def\SmallDet{{\rm D_{50}}}
\def\kgd{\rm kg\mbox{-}days}
\def\pcm2s1{\rm cm^{\mbox{-}2} s^{\mbox{-}1}}
\def\pkkd{\rm \keVee^{\mbox{-}1} kg^{\mbox{-}1} day^{\mbox{-}1}}
\def\MeVee{\rm MeV_{{\rm ee}}}
\def\keVee{\rm keV_{{\rm ee}}}
\def\eVee{\rm eV_{{\rm ee}}}

\def\VVB{{\rm AC^{-} {\otimes} CR^{-} {\otimes} B_{0}}}
\def\VVBc{{\rm AC^{-} {\otimes} CR^{-} {\otimes} B_{c}}}
\def\VTBc{{\rm AC^{+} {\otimes} CR^{-} {\otimes} B_{c}}}

\def\TT{{\rm AC^{+} {\otimes} CR^{+} }}

\def\VV{{\rm AC^{-} {\otimes} CR^{-} }}
\def\VT{{\rm AC^{+} {\otimes} CR^{-} }}
\def\4ctg{{\rm AC^{\pm} {\otimes} CR^{\pm} }}
\def\8ctg{{\rm AC^{\pm} {\otimes} CR^{\pm} {\otimes} B/S}}

\begin{document}

\title{
Studies of Neutrino-Nucleus Elastic Scattering 
with Point-Contact Germanium Detectors 
at the Kuo-Sheng Reactor Neutrino Laboratory
}

\newcommand{\as}{Institute of Physics, Academia Sinica,
Taipei 11529} 
\newcommand{\thu}{Department of Engineering Physics, Tsinghua University, 
Beijing 100084 }
\newcommand{\scu}{College of Physics, 
Sichuan University, Chengdu 610065}
\newcommand{\cusb}{Department of Physics,
School of Physical and Chemical Sciences,
Central University of South Bihar, Gaya 824236}  
\newcommand{\glau}{Department of Physics,
Institute of Applied Sciences and Humanities,
GLA University, Mathura 281406}  
\newcommand{ \hnbgu }{Department of Physics, H.N.B. Garhwal University, 
Srinagar, Uttarakhand 246174} 
\newcommand{\deu}{Department of Physics,
Dokuz Eyl\"{u}l University, Buca, \.{I}zmir 35160} 
\newcommand{\itu}{Department of Physics Engineering,
  Istanbul Technical University, Sarıyer, İstanbul 34467} 
\newcommand{\tku}{Department of Physics, Tamkang University,
  New Taipei City 25137} 
\newcommand{\mrn}{Mirion Technologies (Canberra),
1 chemin de la Roseraie, 67380 Lingolsheim} 
\newcommand{\ndhu}{
Department of Physics, National Dong Hwa University,
Shoufeng, Hualien 97401} 
\newcommand{\ntu}{
Department of Physics, Center for Theoretical Physics, and Leung Center
for Cosmology and Particle Astrophysics, National Taiwan University,
Taipei 10617} 

\newcommand{\corrms}{manu@as.edu.tw}
\newcommand{\corrsk}{skarmakar@as.edu.tw}
\newcommand{\corrgc}{gchandrabhanu@as.edu.tw}
\newcommand{\corrhw}{htwong@as.edu.tw}

\author{ M.K.~Singh }  \altaffiliation[Corresponding Author: ]{ \corrms } \affiliation{ \as } 
\author{ S.~Karmakar } \altaffiliation[Corresponding Author: ]{ \corrsk } \affiliation{ \as } \affiliation{ \glau }
\author{ Greeshma~C. } \altaffiliation[Corresponding Author: ]{ \corrgc } \affiliation{ \as } \affiliation{ \cusb }
\author{ H.B.~Li }  \affiliation{ \as }
\author{ F.K.~Lin } \affiliation{ \as }
\author{ V.~Sharma }   \affiliation{ \as } \affiliation{ \hnbgu }
\author{ L.~Singh }   \affiliation{ \as } \affiliation{ \cusb }
\author{ H.T.~Wong } \altaffiliation[Corresponding Author: ]{ \corrhw } \affiliation{ \as }
\author{ L.T.~Yang } \affiliation{ \thu }
\author{ M.~Agartioglu }  \affiliation{ \as } \affiliation{ \deu } \affiliation{ \ndhu }
\author{ J.H.~Chen } \affiliation{ \as }
\author{ J.W.~Chen } \affiliation{ \ntu }
\author{ C.I.~Chiang } \affiliation{ \as }
\author{ M.~Deniz } \affiliation{ \deu}
\author{ T.~Guo } \affiliation{ \thu} 
\author{ H.C.~Hsu }  \affiliation{ \as }
\author{ W.H.~Kao } \affiliation{ \as } 
\author{ S.~Karada\v{g} } \affiliation{ \as } \affiliation{ \itu } \affiliation{ \tku }
\author{ J.B.~Legras } \affiliation{ \mrn }
\author{ C.H.~Leung }  \affiliation{ \as }
\author{ J.~Li }  \affiliation{ \thu }
\author{ T.Y.~Liang }  \affiliation{ \as }
\author{ S.T.~Lin } \affiliation{ \scu }
\author{ S.K.~Liu } \affiliation{ \scu }
\author{ H.~Ma }  \affiliation{ \thu }
\author{ V.~Marian } \affiliation{ \mrn }
\author{ D.~Mishra }  \affiliation{ \as } \affiliation{ \cusb } 
\author{ M.K.~Pandey } \affiliation{ \ntu }
\author{ K.~Rani }  \affiliation{ \cusb } 
\author{ K.~Saraswat } \affiliation{ \as } 
\author{ M.K.~Singh } \affiliation{ \as }  \affiliation{ \glau} 
\author{ V.~Singh } \affiliation{ \cusb} 
\author{ D.~Tanabe } \affiliation{ \as }
\author{ J.S.~Wang } \affiliation{ \as }
\author{ Y.F.~Wang } \affiliation{ \thu} 
\author{ C.-P.~Wu } \affiliation{ \ndhu }
\author{ C.H.~Yeh }  \affiliation{ \as }
\author{ Q.~Yue } \affiliation{ \thu }

\collaboration{ TEXONO Collaboration }

\date{\today}

\begin{abstract}
  The low energy and intense flux of electron anti-neutrinos
  from nuclear reactors provide the perfect stage to study
  elastic neutrino-nucleus scattering ($\nuAel$) in the fully
  coherent regime. We report results from the TEXONO experiment
  using electro-cooled $p$-type point-contact Germanium detectors
  with masses of 523~g and 1434~g at the Kuo-Sheng Reactor
  Neutrino Laboratory. We report improved constraints on the
  $\nuAel$ cross section with a combined exposure of
  404(813.7)~$\kgd$ of Reactor ON(OFF) data at an
  electron-equivalent threshold of 200~$\eVee$. The Lindhard model,
  in which the quenching factor is parameterized by a single
  parameter $k$, is adopted to describe the suppression of
  ionization yield. At the benchmark value of $k{=}$0.162, a
  limit of $\rho<$2.0 at 90\% confidence level (CL) is
  derived, where $\rho$ represents the ratio of the observed
  to the predicted Standard Model cross section. Moreover,
  the region $k>$0.205 is excluded at 90\% CL using the
  SM-predicted $\nuAel$ rate. A bound on the neutrino magnetic
  moment from $\nuAel$ at $\munu {<} 5.9\times10^{-10}~\muB$
  at 90\% CL is also derived.
  
\end{abstract}

\maketitle

\section{Introduction}

Tremendous progresses were made in the discovery and investigations
of neutrino masses in mixings in the past three decades.
Nevertheless, answers to several fundamental questions remain
elusive, such as: What are the exact structures and absolute
scales of the neutrino masses? Do they undergo ``Beyond the Standard Model''
(BSM) interactions? And whether neutrinos are
their own antiparticles with identical CP
properties~\cite{PDG2024-neutrino}? Neutrino
interactions with matter play important roles in these
studies, both as tools for investigation and as probes
for exploration.

\begin{figure}
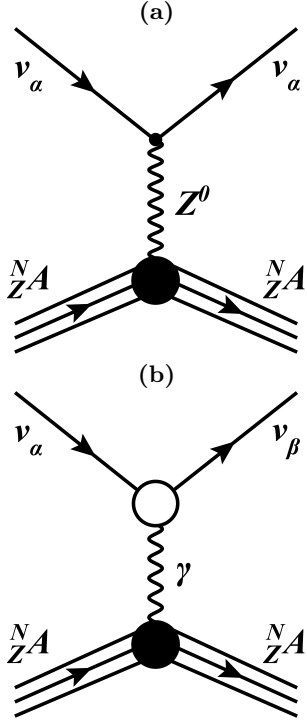

  \begin{center}
  {\bf (a)}\\
  \includegraphics[width=4.0cm]{fig1a.pdf} \\
  {\bf (b)}\\
  \includegraphics[width=4.0cm]{fig1b.pdf} 
  \end{center}
  \caption{Feynman diagrams of $\nuAel$ in the 
    (a) SM and (b) BSM-$\munu$ channels.
    Interactions involving $\munu$ induce a change in
    helicity states between initial-state ($\nu_{\alpha}$)
    and final-state ($\nu_{\beta}$) neutrinos, and
    therefore do not interfere with the SM electroweak
    interactions coupled by the neutral weak
    boson $Z^{0}$.
  }
  \label{fig::nuA-Feynman}
\end{figure}

The coherent elastic scattering of neutrinos
with the nucleus $A(Z,N)$
($\nuAel$, also denoted as CE$\nu$NS in 
the literature)~\cite{DZFreedman:1974,Freedman:1977,ARNPS2023}:
\begin{equation}
 \nu  ~ {+} ~  A(Z,N) ~  {\rightarrow} ~  \nu ~  {+} ~  A(Z,N) ~~ ,
\label{eq::nuAel}
\end{equation}
as depicted in Figure~\ref{fig::nuA-Feynman}(a),
is a fundamental electroweak neutral-current process
in the Standard Model (SM). 
It has large cross section relative to other neutrino interactions,
due to enhancement by quantum-mechanical coherency effects~\cite{Kerman:2016jqp},
at low $\q2$, where
$ q \equiv | \vec{q} |$, the three-momentum transfer.


This article reports experimental studies
of $\nuAel$ with Electro-cooled $p$-type Point-Contact
Germanium (Ge) detectors (ECPCGe)
at the Kuo-Sheng Reactor Neutrino Laboratory (KSNL)
as part of the TEXONO research program~\cite{TEXONO:2018}.
The analysis and results are based on 404(813.7)~$\kgd$
of Reactor ON(OFF) physics data, significantly
expanding from our earlier publication with 242(357)~$\kgd$
of Reactor ON(OFF) exposure~\cite{TEXONO-PRL2025}.
In addition to the increased data size, the
improvement in sensitivity is also attributed to the
inclusion of high energy tail of the reactor electron
anti-neutrinos ($\nuebar$)
spectrum~\cite{PRL:2022:DayB,CONUS-2}. Experimental
details and various supporting measurements are also
presented.

The paper is structured as follows. 
Section~\ref{sect::nuAel} summarizes the history and
status of the experimental studies of $\nuAel$, both
as a SM interaction and as a probe of BSM processes, using
neutrino magnetic moments ($\mu_{\nu}$) as an example.
The roles of Ge-detectors in $\nuAel$ studies, as well as the 
evolution to ECPCGe, are described in Section~\ref{sect::ECPCGe}.
This is followed by discussions in Section~\ref{sect::TEXONO} on
the conceptual design, experimental configurations, as well as
the signal selection procedures in data analysis.
The measurements and scientific results are presented in
Section~\ref{sect::results}. The energy transferred in $\nuAel$
to the target nuclei is via nuclear recoils with kinetic
energy $\TNR$. The measured energy response of ECPCGe,
characterized by the quenching factor (QF), is incorporated
to convert the measured electron-recoil energy $T$. The
electron-equivalent unit $\eVee$ is adopted throughout
the article to describe the measured energy, unless otherwise
stated.

\section{Neutrino-Nucleus Elastic Scattering}
\label{sect::nuAel}

Proposals in the 2000s for experimental studies of
$\nuAel$ with reactor $\nuebar$ using low-threshold
Ge-detectors~\cite{Wong:2005vg} triggered
intense experimental efforts~\cite{ARNPS2023,Wong2025}.
Proposals employing decay-at-rest pions ($\DARpi$) provided by
spallation neutron source soon followed~\cite{scholberg}.
First measurement of $\nuAel$ was achieved with
$\DARpi$-$\nu$ in the COHERENT experiment with
CsI(Na) scintillator~\cite{COHERENT_SCIENCE:2017},
followed by measurements with
liquid Ar~\cite{COHERENT:LAr:PRL2021}
and Ge-detectors~\cite{COHERENT:Ge:Adamski:2024yqt}.
The $\nuAel$ events from solar and atmospheric neutrinos are
the irreducible ``neutrino fog'' background
in dark matter (DM) experiments~\cite{PDG2024-DM}.
Positive signatures of $\nuAel$ from $^{8}$B solar neutrinos
have recently been observed~\cite{PandaX-4T:PRL2024,
  XENONnT:PRL2024,XENONnT:2026::8B}, which can probe into
the ``neutrino fog'', an irreducible background relevant
for direct DM searches. The CONUS+ experiment
has recently reported a positive signature of reactor
$\nuAel$ with ECPCGe at a 3.7$\sigma$
significance~\cite{CONUS+2025-2,CONUS+2025-1}.

The $\nuAel$ process of Eq.~(\ref{eq::nuAel})
and Figure~\ref{fig::nuA-Feynman}(a)~\cite{Papoulias:2015vxa,nuA-nFF-2,Cadeddu_2023}
can provide sensitive tests of the SM electroweak 
interactions~\cite{nuA-SM-1,nuA-SM-2}
and serve as a probe of 
BSM physics~\cite{nuA-BSM-1,nuA-BSM-2,MDeniz:PRD2025,Karmakar:INJP2025}
at low $\q2$.
It may open new avenues for the study of
neutron density distributions~\cite{nuA-nFF-1,nuA-nFF-2},
supernova neutrino detection~\cite{nuA-SNnu-1,nuA-SNnu-2}, and
real-time monitoring of nuclear reactors
using compact and transportable neutrino 
detectors~\cite{nuA-Rmon-1,nuA-Rmon-2,nuA-Rmon-3}.

The $\nuAel$ reaction provides a unique laboratory
to study the quantum-mechanical coherency effects 
in electroweak interactions. At low momentum transfer,
the de Broglie wavelength of the neutrino is large
compared with the nuclear size, such that the
scattering amplitudes from individual nucleons add
coherently, leading to an enhanced cross section.
The neutrino energy ($\Enu$)
and $\TNR$ are typically much smaller than the
target nucleus mass ($\MN$).
The corresponding value of $\q2$ is determined
by the scattering kinematics as
\begin{equation}
\q2 ~ = ~ 2 \MN \TNR ~ + ~ \TNR^2 \approx 2 \MN \TNR  ~~ .
\end{equation}
Kinematics constrains the maximum
recoil energy to be 
\begin{equation}
\TNR^{max} ~ = ~
2 \Enu ^2 / ( M + 2 \Enu ) \approx  2 \Enu ^2 / M  ~~ .
\label{max::reco}
\end{equation}
A generic scale of $\Enu$$<$50~MeV  
is usually taken to characterize the 
requirement of coherency.
The degree of coherency of the $\nuAel$ interactions 
was parametrized and quantified
in Ref.~\cite{Kerman:2016jqp}.
 
\subsection{Standard Model Weak Interaction} 

The first milestone in studies of $\nuAel$ is
the observation and measurement of the SM cross section.
The differential cross section of $\nuAel$ in the
SM is given by~\cite{Papoulias:2015vxa}:
\begin{eqnarray}
\label{eq::diffsigmanuA}
\frac{d \sigmanuA}{d \TNR} ~ &  =  & ~    
 2 \MN  ~ \left[ \frac{{\it d} \sigmanuA}{{\it d q}^{2}} \right] \\ \nonumber
&   =   &  
2 \MN  ~ \Bigg[ ~  \frac{1}{2} ~ \left( \frac{ G_F^2 }{ 4 \pi}  \right) \cdot 
\left( 1 - \frac{ {\it q}^{2} }{ 4 \Enu ^2 } \right) \cdot \\ \nonumber & &   
\hspace*{1cm} \Big( \varepsilon Z F_Z ( \q2 ) - N F_N ( \q2 ) \Big) ^2 ~ \Bigg]   ~~,
\end{eqnarray}
where $F_Z ( \q2 )$ and $F_N ( \q2 )$ are the
proton and neutron nuclear form factors, respectively,
for a nucleus $A(Z,N)$, while
$\varepsilon$$\equiv$(1$- 4 ~ {\rm sin^2 \theta_W } )$=0.045,
where ${\rm sin^2 \theta_W }$ is taken at low $\q2$,
indicating that the dominant contributions
arise from neutrons.

The total cross section depends on
$( \Enu , \T0 ;  M , Z , N )$
and is given by:
\begin{equation}
\sigmanuA = 
\int_{\q2_{min}}^{\q2_{max}} 
\left[ \frac{d \sigmanuA }{d \q2} ( \q2 , \Enu )  \right] ~ d \q2 ~~,
\label{eq::totalsigmanuA}
\end{equation}
where $\T0$ is the detector threshold, while the integration limits of
$\q2_{max}$=$4 \Enu ^2 [ M / ( M$+$2 \Enu ) ]$$\approx$$4 \Enu ^2$ and
$\q2_{min}$=$2 M \T0$ are defined by
the kinematics and detection threshold, respectively.
A detector threshold of $\T0 {<} 2 \Enu ^2 / M$ 
is required to detect
neutrinos of energy $\Enu$.
In this work, we follow the convention of
expressing the sensitivities with the
parameter $\rho$, which is the ratio of the
measured experimental cross section to the
SM cross section $\sigmanuA$.

We adopt the effective method of
  Refs.~\cite{Engel::PLB1991,Helm::PR1956},
which assumes identical form factors for neutrons
and protons:
$F_{Z} ( \q2 )$=$F_{N} ( \q2 )$$\equiv$$F ( \q2 )$$\in$[0,1],
with
\begin{equation}
F ( \q2 ) =
\left[ \frac{3}{q R_{0}} \right] ~
J_{1} ( q R_{0} )
~ {\rm exp}  \left[ - \frac{1}{2} \q2 s^2 \right]  ~~~ ,
\label{eq::formfactor}
\end{equation}
where $J_{1} ( x )$ is
the first-order spherical Bessel function.
The target nuclear dependence is introduced through
$R_{0}^2$=$R^2$$-$$5 s^2$, with $s$=0.5~fm and
$R$=$1.2 A^{\frac{1}{3}} ~ {\rm fm}$.
This scheme is widely adopted in other $\nuAel$-related
studies~\cite{CONUS+2025-1,VS:2021PRD,Kerman:2016jqp,Corona:PRD2025nuNCh,TEXONO-PRL2025}, 
so that this choice facilitates direct
cross-comparison of results. An alternative
derivation~\cite{Patton::PRC2012} yields
form factors consistent to within $<$0.07(1.7)\%
over the kinematic ranges
corresponding to $\Enu$=10(50)~MeV.

There are intense experimental programs on
$\nuAel$~\cite{ARNPS2023}, with neutrinos
from reactors or from
$\DARpi$-$\nu$ sources~\cite{scholberg}. The coherency
for $\DARpi$-$\nu$ is partial, typically in the range
of 50$-$80\%~\cite{Kerman:2016jqp,VS:2021PRD} for medium-to-heavy
target nuclei. On the contrary, coherency is nearly complete 
for reactor $\nuAel$ at $>$99\% (equivalently,
$F_Z(\q2)\approx1$ at $\q2\rightarrow$0),
therefore providing excellent prospects for
precision SM measurements and 
enhanced sensitivity for BSM physics searches. 
However, $\TNR$ is typically below a few keV,
since $\Enu {<}  11 ~ {\rm MeV}$ for reactor $\nuebar$.
Both the low observable energy and
the low event rates pose formidable challenges for
experimental studies of reactor $\nuAel$.

\subsection{Beyond Standard Model Neutrino Magnetic Moments}

Neutrino-photon couplings arise as a generic
consequence of neutrino mass and constitute
intrinsic neutrino properties BSM~\cite{Giunti:2024gec}.
Among these, $\munu$  provides a particularly sensitive probe
in low energy experiments~\cite{TEXONO:2007xds}, including those on $\nuAel$. 
The $\nuAel$ process induced by $\munu$ is
coherent ($F_{\rm em} (q^{2}\rightarrow$0){$\approx$}1),
and the cross section scales as $Z^{2}$.
There is no interference with the SM
$Z^{0}$-exchange process, since it involves a
change in helicity between the initial
($\nu_{\alpha}$) and final-state neutrinos
($\nu_{\beta}$), as shown in
Figure~\ref{fig::nuA-Feynman}(b). The $\munu$-induced
differential cross section for odd-$A$ (e.g. $^{73}$Ge)
and spin-zero nuclei (e.g. $^{70,72,74,76}$Ge) are given
by~\cite{PRD::1989::PVEJ}, respectively:
\begin{equation}
  \begin{aligned}
    \Bigg[\frac{d\sigmanuA}{d\TNR}\Bigg]_{{\rm odd}\text{-}A} &= \frac{\pi \alpha^{2} \munu^{2}}{m_{e}^{2}}
    \Bigg[
      \frac{1 - \TNR/\Enu}{\TNR} Z^{2}
      - \frac{\TNR}{2\Enu^{2}} \mu_{N} Z \\
      &\quad + \frac{(2 - \TNR/\Enu)^{2} - 2M\TNR/\Enu^{2}}{8M} \mu_{N}^{2}
      \Bigg],~~ {\rm and}
    \label{eq:A6}
  \end{aligned}
\end{equation}

\begin{equation}
  \Bigg[\frac{d\sigmanuA}{d\TNR}\Bigg]_{\rm spin\text{-}zero} = \frac{\pi \alpha^{2} \munu^{2}}{m_{e}^{2}}
  \Bigg[
    \frac{1 - \TNR/\Enu}{\TNR} 
    + \frac{\TNR}{4\Enu^{2}}
    \Bigg]Z^{2},
  \label{eq:A7}
\end{equation}
where $m_{e}$ is the electron mass, $\alpha$ is
the fine-structure constant, and $F_{\rm em} (q^{2})$ is
the nuclear electromagnetic form factor. The anomalous
magnetic moment is expressed in units of the nuclear
magneton $\mu_{N}$. In the low recoil energy regime
($\TNR$/$\Enu\ll$~1), where $F_{\rm em} (q^{2}\rightarrow$0){$\approx$}1,
the cross section is simplified to:
\begin{equation}
  \frac{d\sigmanuA}{d\TNR} \approx \frac{\pi \alpha^{2} \munu^{2}}{m_e^{2}} \frac{Z^{2}}{\TNR},
  \label{eq:A8}
\end{equation}
and can be characterized by enhancement at low
energy via the $1/\TNR$ dependence.

Reactor-based $\nuAel$ experiments,
with $\nuebar$ as the source and detectors with
sub-keV sensitivity, offer strong potential
to probe $\munu$.
Existing limits~\cite{MUNU::2005plb,GEMMA::2012nuEM,
  Krasnoyarsk::Vidya:1992nf,Derbin:1993::Rovno,
  TEXONO:2007xds,CONUS:2022qbb::HBone,
  CONUS:2026uhz:nu:EM} are primarily derived from the
$\nuebar$-electron scattering channel, with the
most stringent constraint, $\munu {<} 2.9 {\times} 10^{-11} ~ \muB$
at 90\% CL, reported by the GEMMA
experiment~\cite{GEMMA::2012nuEM}, where $\muB$
denotes the Bohr magneton.
An analysis~\cite{Corona:PRD2025nuNCh} on the
$\nuAel$ channel using data from the
TEXONO~\cite{TEXONO-PRL2025} and
CONUS+~\cite{CONUS+2025-1} experiments was
recently reported.

\section{Reactor Neutrino with Germanium Detectors}
\label{sect::ECPCGe}

\subsection{Development and Evolution}

The TEXONO program at KSNL~\cite{TEXONO:2018}
pioneered the use of high-purity
Ge ionization detectors (HPGe) in studies
of $\munu$ with reactor
$\nuebar$~\cite{TEXONO:2002pra,PRD::2005texono,
  TEXONO:2007xds}, achieving a reduction in
the detector energy threshold from
$\mathcal{O}$(1~MeV$_{\rm ee}$) to
$\mathcal{O}$(10~$\keVee$). This advancement
triggered proposals to observe $\nuAel$ at
reactors using Ge-detectors with further
reduced thresholds of $\mathcal{O}$(100~$\eVee$).
This inspired the first realization of a 
$p$-type PCGe detector with 0.475~kg mass and
$\approx$330~$\eVee$ threshold by the
CoGeNT program~\cite{Barbeau::JCAP2007,
  CoGeNT:2008yoi}, and was further
consolidated through decades of R\&D efforts
across several generations of progressively
improving PCGe technologies~\cite{TEXONO:2016:AKSoma}.
The availability of the sub-keV detection
window set the stage for new studies at KSNL,
including investigations of neutrino
millicharge~\cite{Chen:2014dsa} and various
DM scenarios~\cite{TEXONO-LDM2009,
  TEXONO-LDM2013,LSingh_PRD2019,
  MKSingh_CJP2019}. It also inspired
{\it ab initio} atomic ionization
calculations~\cite{Texono:Th:CHEN:2014} and
studies of atomic effects in neutrino
electromagnetic interactions with
Ge-detectors~\cite{JWChen:PRD::2015nEM}.
These developments further catalyzed the
construction of the China Jinping
Underground Laboratory
(CJPL)~\cite{Cheng:2017usi} and the
CDEX DM program~\cite{PRD::CDEX::2014}. This
line of research on reactor $\nuebar$
with sub-keV detector sensitivity
will be continued at the Sanmen
Reactor Neutrino Laboratory, currently
under construction in Zhejiang,
China~\cite{Yang:20249T,RECODE::SChina::2025,
  Dai:2025kai}.

\subsection{Reactor $\nuAel$ }

Reactor $\nuebar$, with its high flux and low energy,
provides an excellent laboratory to study $\nuAel$.
It can probe quantum-mechanical coherency in
weak neutral-current interactions in the near-complete 
regime~\cite{Kerman:2016jqp,VS:2021PRD}, where the
cross section scales approximately as $N^2$. This
suppresses ambiguities and degeneracies associated
with $\nuAel$ at $\DARpi$-$\nu$, and opens windows
for precision measurements of SM
cross sections~\cite{Cadeddu_2023,DeRomeri2023CEnuNS,Annurev:Xu::2023}
and BSM searches~\cite{PANDEY::2024::PPNP,nuA-BSM-1,nuA-BSM-2}.

There are intense efforts to study reactor
$\nuAel$ in current and next-generation experiments.
The completed PCGe programs include
TEXONO~\cite{TEXONO-PRL2025,Wong2025}, CONUS~\cite{Bonet:2020awv,CONUS-2},
and DRESDEN~\cite{DRESDEN-2,DRESDEN-1}, while
the active ones are $\nu$GeN~\cite{nuGeN-2,nuGeN-1} and
CONUS+~\cite{CONUS+2025-2,CONUS+2025-1},
to be joined by RECODE~\cite{Yang:20249T,Dai:2025kai}
in the future. The key experimental features are
summarized in Table~\ref{tab::expt}. Among the
reactor PCGe experiments, the DRESDEN experiment
reported positive signatures~\cite{DRESDEN-1}, but
these were subsequently refuted by other
experiments~\cite{CONUS-2,TEXONO-PRL2025,nuGeN-1}.
CONUS+ recently reported a 3.7$\sigma$ positive
observation~\cite{CONUS+2025-1}. The final results of
the TEXONO $\nuAel$ program are presented in this
work.

Reactor $\nuAel$ projects employing alternative
detector technologies include CONNIE with Si-Skipper
CCDs~\cite{CONNIE:PRD2019}; MINER~\cite{MINER:arXiv:2017},
NUCLEUS~\cite{NUCLEUS:PRL2023}, and
RICOCHET~\cite{RICOCHET:arXiv2023}
with cryogenic bolometers; 
as well as RED-100 which is based on
both liquid argon~\cite{RED::LAr::2023}
and xenon~\cite{RED-100:2024izi,RED:JINST2013} techniques.


\begin{figure}
  \begin{center}
    \includegraphics[width=8.2cm]{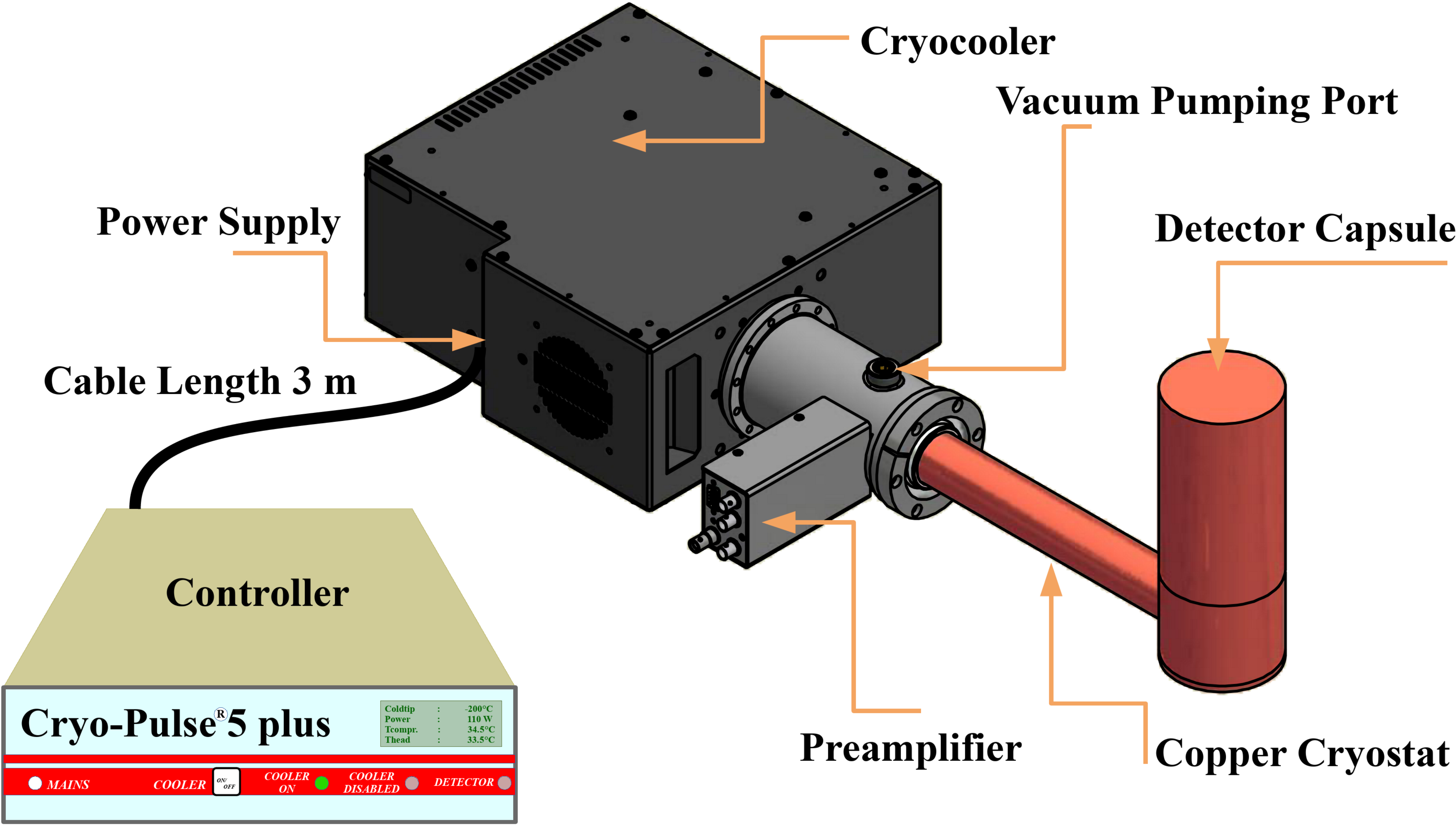} 
  \end{center}
  \caption{Schematic diagram of the ECPCGe detector.
  }
  \label{fig::ECGe}
\end{figure}


\begin{table*}
  \centering
  \caption{Summary of the key parameters and results of
    reactor $\nuAel$ experiments using PCGe detectors.}
\begin{center} 
\renewcommand{\arraystretch}{1.2} 
\begin{tabular}{|l|c|c|c|c|c|c|c|c| } 
\hline 
Experiment & Reactor     & Core thermal   & Distance  & $\phi ( \nuebar )$  & 
~~ Target ~~ & $\T0$     & ON background & $\rho$ (at $k$=0.162)         \\ 
           & plant       & power          & from core &  ($\pcm2s1$)        & 
mass   & ~~ ($\eVee$) ~~ & at threshold        & $\left[  {\rm Limits~at~90\%~CL } \right]$           \\ 
           &             & (GW$_{\rm{th}}$) & (m)       &                     & (kg)   &           & ($\pkkd$)          &                            \\ \hline \hline 
DRESDEN    & Dresden-II  & 2.96           & 10.39     &  48$\times10^{12}$   & 2.924  & 200       & $\sim$3100         & $<$ 7.32~\cite{DRESDEN_PRL}    \\ 
$\nu$GeN   & Kalinin     & 3.1            & 11.1      &  44$\times10^{12}$   & 1.41   & 290       & $\sim$43           & $<$ 4.3~\cite{nuGeN-1}         \\
CONUS      & Brokdorf    & 3.9            & 17.1      &  23$\times10^{12}$   & 3.73   & 210       & $\sim$16           & $<$ 1.57~\cite{CONUS-2}        \\ 
CONUS+     & Leibstadt   & 3.6            & 20.7      &  15$\times10^{12}$   & 2.83   & 160       
& $\sim$250          & $1.14 \pm 0.36$~\cite{CONUS+2025-1}   \\ \hline
\multirow{3}{*}{TEXONO}  & \multirow{3}{*}{KSNL} & \multirow{3}{*}{2.9} & \multirow{3}{*}{28} &  
\multirow{3}{*}{6.37$\times10^{12}$} & 1.434 & \multirow{3}{*}{200} & 112 & 
$<$ 4.7 ~\cite{TEXONO-PRL2025}  \\
           &&&&& $\BigDet :$ 1.434 & & \multirow{2}{*}{85} & \multirow{2}{*}{$<$ 2.0 [This work]}       \\    
& & & & & $\SmallDet :$ 0.523 & &  & \\ \hline  
RECODE     & \multirow{2}{*}{Sanmen}& \multirow{2}{*}{3.4} & 11 (Near) & \multirow{2}{*}{$\mathcal{O}(10^{13})$} & \multirow{2}{*}{10}& \multirow{2}{*}{160} & \multirow{2}{*}{2} & In preparation  \\ 
(Proposal) &             &                & 22 (Far)  &&&&&~\cite{Dai:2025kai,Yang:20249T,Shen:2026kzy} \\ \hline

\end{tabular} 
\end{center} 
\label{tab::expt} 
\end{table*} 


\subsection{Point-Contact Germanium Detector} 
\label{pcge::ecge}

With their low energy threshold, excellent
energy resolution, and high intrinsic radiopurity,
Ge ionization detectors have opened
new avenues and contributed significantly in
the studies of low energy neutrino physics, light
DM searches, and neutrinoless double-beta
decay~\cite{AVIG::FT::FR::2019,TEXONO:2016:AKSoma}.
A summary of PCGe-based experiments for $\nuAel$
is given in Table~\ref{tab::expt}.

The TEXONO program has adopted HPGe and
PCGe detectors over several generations
of experiments at KSNL~\cite{TEXONO:2007xds,
  TEXONO-LDM2013,LSingh_PRD2019}. Liquid
nitrogen in dewars was used as a coolant,
and regular intervention was required.
Recent advances in electro-cooling
technology compatible with low-noise
applications have lead to the realization of
ECPCGe, in which the
necessity of maintenance and intervention
is significantly reduced. This merit turned
out to be essential to allow the data taking
during the COVID-19 pandemic periods, when
human access to the KSNL facilities was
prohibited or severely restricted.

The ECPCGe detectors~\cite{Mirion:2020}
used in this work were custom-designed and
built. The cooling technology is with
Cryo-Pulse\textsuperscript{\textregistered}~5~Plus
(CP5+) cryostat\footnote{Mirion
Technologies, formerly Canberra.}.
The integrated cryocooler and detector are
depicted schematically in
Figure~\ref{fig::ECGe}. Radiopure
electrolytic copper is used for the
cryostats, which are thermally coupled to the PCGe
diodes via cooling arms of approximately 20~cm in length.
The CP5+ cryocooler is a Stirling pulse-tube
refrigerator based on the Linear Pulse
Tube series\footnote{LPT9310 pulse-tube
cryocooler (Thales Cryogenics).},
employing a two-stage
design to achieve adjustable temperatures of
approximately $-$190~to~$-$200\textdegree C~\cite{deWaele:2011,Willems:2015}.
A hermetically sealed, non-flammable, non-CFC
refrigerant is cyclically compressed and
expanded, generating pressure
oscillations~\cite{Mirion:2020,deWaele:2011}
that create a temperature gradient along
the pulse-tube without moving parts in the
cold region, thus minimizing vibration and
mechanical wear~\cite{PAN2020103096,PAN2020103097}.
Heat is extracted at the pulse-tube cold
tip, which is thermally coupled to the PCGe
diode, while the warm end dissipates heat to the
ambient environment. The Ge-crystal sensor
effectively attains the cold-tip temperature,
with a residual offset of
$\pm$0.05\textdegree C after CP5+ power
stabilization. The sealed, maintenance-free
design ensures high reliability and long-term
operation~\cite{Mirion2022}, enabling
continuous data acquisition with minimal maintenance.
Additional technical details on the
cryocooler and cooling technology employed
can be found in Refs.~\cite{Hakenmuller2020,Bonet:2020ntx}.


\begin{figure}
  \begin{center}
    {\bf (a)}\\
    \includegraphics[width=8.0cm]{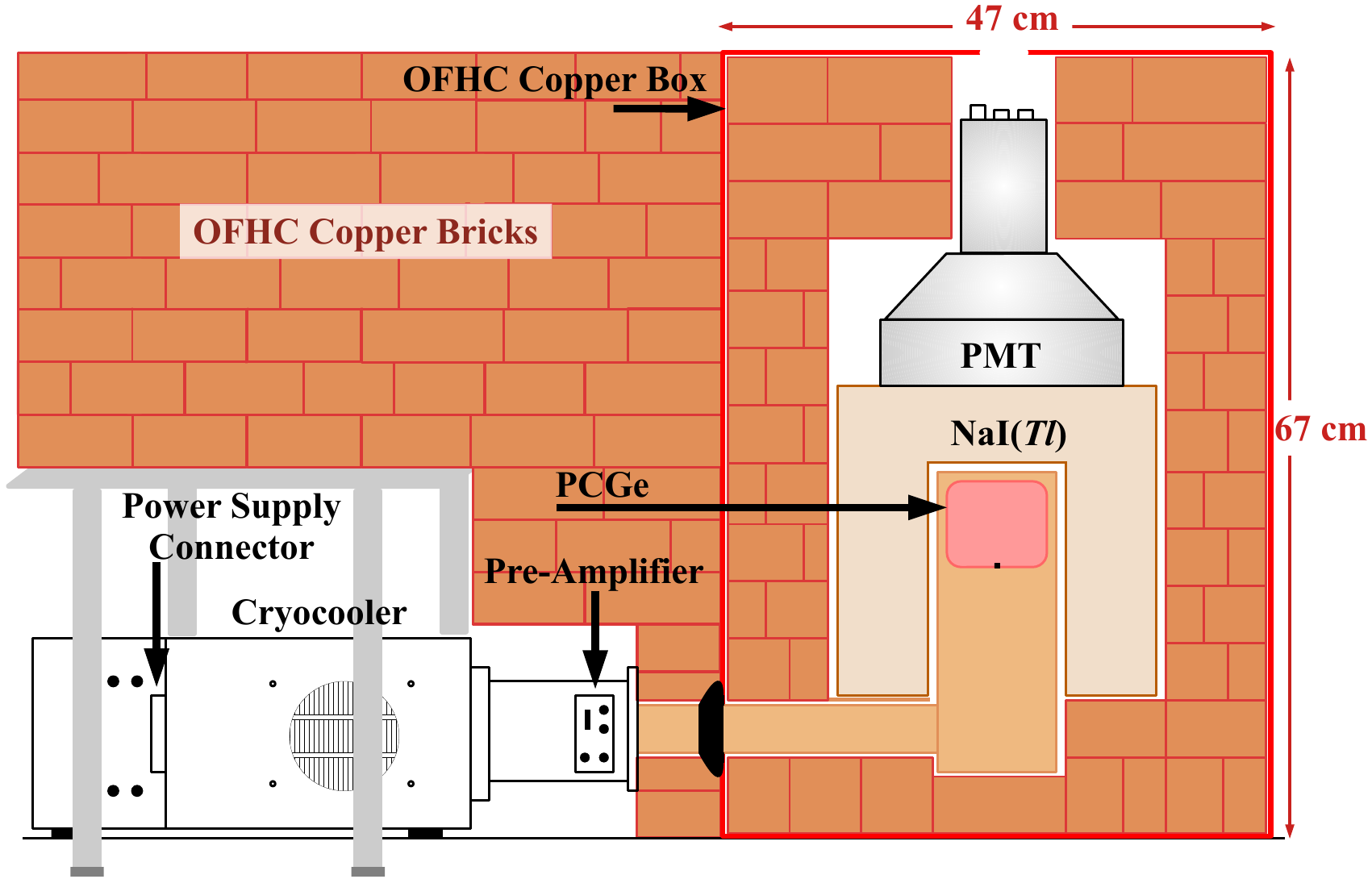} \\
    {\bf (b)}\\
    \includegraphics[width=8.0cm]{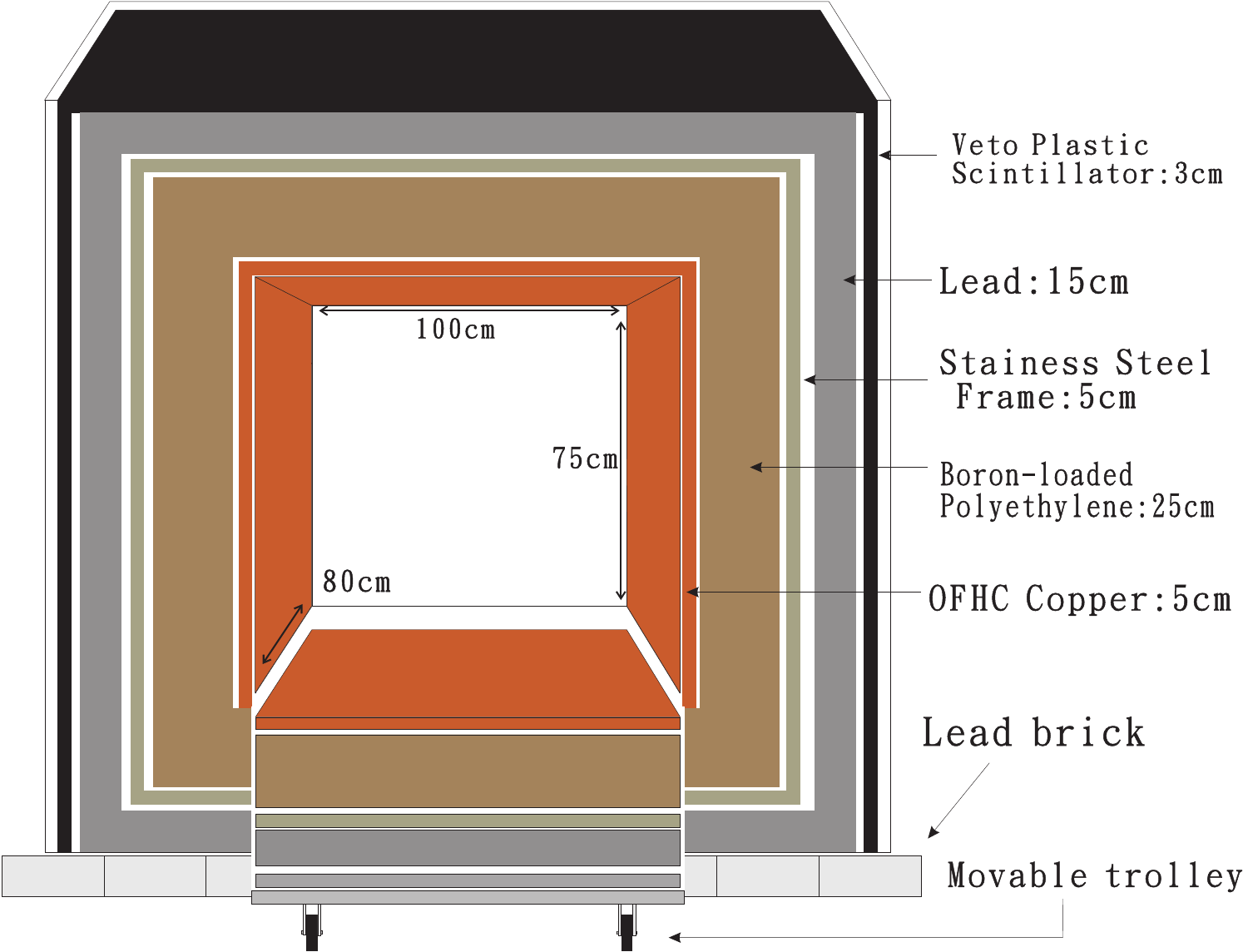} \\
    {\bf (c)}\\
    \includegraphics[width=8.0cm]{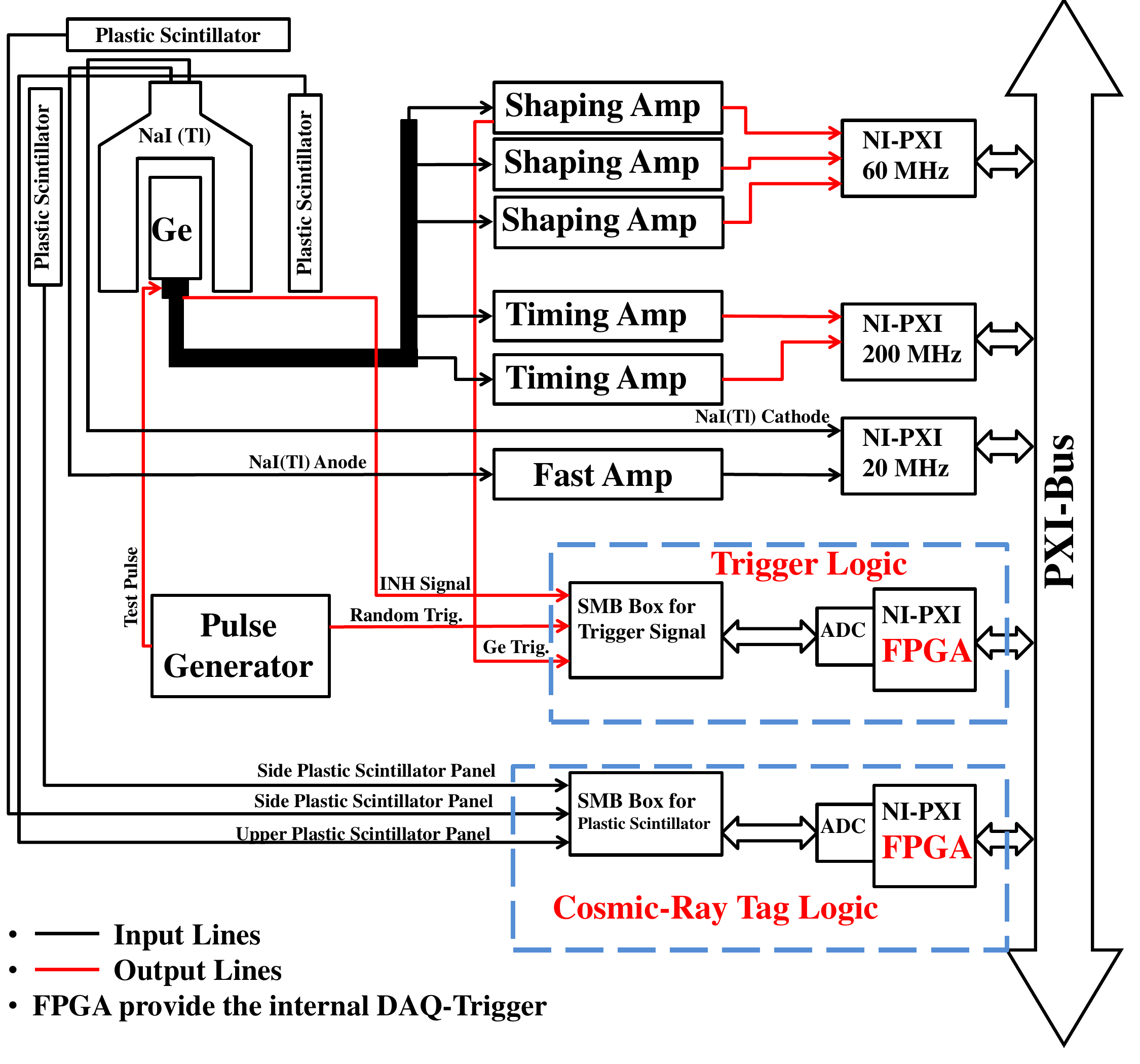} 
  \end{center}
  \caption{Schematic configurations of the TEXONO
    $\nuAel$ experiment at KSNL:
    (a) the ECPCGe and NaI(Tl) AC detectors, together
    with the inner shielding composed of OFHC copper bricks; (b)
    the shielding structure with CR veto panels,
    inside which the experiment was located; and (c)
    the electronic readout and data acquisition
    systems. 
  }
  \label{fig::expt}
\end{figure}

\section{The TEXONO Experiment}
\label{sect::TEXONO}
\subsection{Experimental Configurations}

The conceptual design of the $\nuAel$
experiment follows that of previous
experiments using Ge-detectors at
KSNL~\cite{TEXONO:2007xds,TEXONO-LDM2009,
  TEXONO-LDM2013,LSingh_PRD2019}. A
schematic diagram of the experimental
setup is depicted in Figure~\ref{fig::expt}(a).
The main difference is that the conventional
liquid nitrogen dewar is replaced by
the electro-cooling technology in the
ECPCGe system.

Data taken with two ECPCGe detectors
featuring different Ge-sensor masses, were used in
this analysis: (1) a 1434~g detector with 70~mm
diameter and 70~mm height ($\BigDet$),
and (2) a 523~g detector with 50~mm diameter and
50~mm height ($\SmallDet$). Both ECPCGe detectors
are enclosed within well-shaped
NaI(Tl) crystal scintillators, which serve as
Anti-Compton (AC) detectors. This analysis
corresponds to total Reactor ON(OFF) valid
exposures of 404(813.7)~$\kgd$, obtained
by combining the $\BigDet$ and $\SmallDet$
datasets. A summary of the data taking periods
is given in Table~\ref{tab::ksdaq}.

The ECPCGe detectors, NaI(Tl)-AC detectors, and passive
shielding are housed within a 50-ton shielding
structure, schematically depicted in
Figure~\ref{fig::expt}(b). It consists of,
from outside to inside, 2.5~cm thick plastic
scintillator panels with photomultiplier
tubes (PMTs) readout for cosmic-ray (CR)
veto, 15~cm of lead, 5~cm of stainless
steel support structures, 25~cm of
boron-loaded polyethylene, and 5~cm of
oxygen free high conductivity (OFHC)
copper. The innermost volume, with
dimensions of 100$\times$80$\times$75~cm$^{3}$, provides
the flexibility to accommodate different
detectors for diverse physics studies.

\subsection{Data Taking with ECPCGe}
A schematic diagram of the electronics readout and
data acquisition (DAQ) systems~\cite{TEXONO:2016:AKSoma}
is illustrated in Figure~\ref{fig::expt}(c). Signals
from Ge-crystal sensors are first amplified by
front-end JFETs\footnote{Custom-built for low
electronics noise, Mirion Technologies Lingolsheim.
Contact the company for technical information.}
located in the vicinity of the PCGe diodes. The outputs
are then fed to reset preamplifiers\footnote{Model PSC957
(D$_{70}$) and PSC854P (D$_{50}$) Mirion Technologies
Lingolsheim.} positioned $\sim$30~cm away.

The preamplifier signals are
further processed by both shaping amplifiers (SA) and timing
amplifiers (TA)\footnote{Canberra 2026 and 2111,
respectively.}. Output from the TA preserves
rise-time ($\tau$) information for
distinguishing bulk ($B$) and surface ($S$)
events, to be further discussed in
Section~\ref{sect::signalselection}. The SA
signals, with a shaping time of 6~$\mu$s  
are optimized for energy measurement. These SA
signals are fed to a real-time FPGA-based
discriminator\footnote{National Instruments
PXIe-7961R FlexRIO $-$ FPGA}, whose output
provides the physics trigger instant for the
DAQ. Pulses from the TA and SA are digitized
by 200~MHz and 60~MHz flash analog-to-digital
converters\footnote{National Instruments PXI
5124 and PXI 5105, respectively.}, respectively.
Multi-channels with different SA gain settings
allow high energy events to be recorded. The
DAQ dynamic range is $\sim$(0.1-467)~$\keVee$
and (0.1-900)~$\keVee$ for $\BigDet$ and
$\SmallDet$, respectively. Signals from the AC and
CR detectors are also recorded once the
triggers are generated.

In addition to the physics triggers,
random trigger (RT) events are
recorded at a rate of 0.1~Hz to monitor system
noise behavior and to provide efficiency
measurements. The DAQ system also includes a
programmable analog test pulser\footnote{National
Instruments PXI~5412.}, which can generate
pulse shapes matching ECPCGe signals
for the purposes of energy calibration
and pulse shape analysis.

\begin{figure}
  \begin{center}
  {\bf (a)} 
  \includegraphics[width=8.2cm]{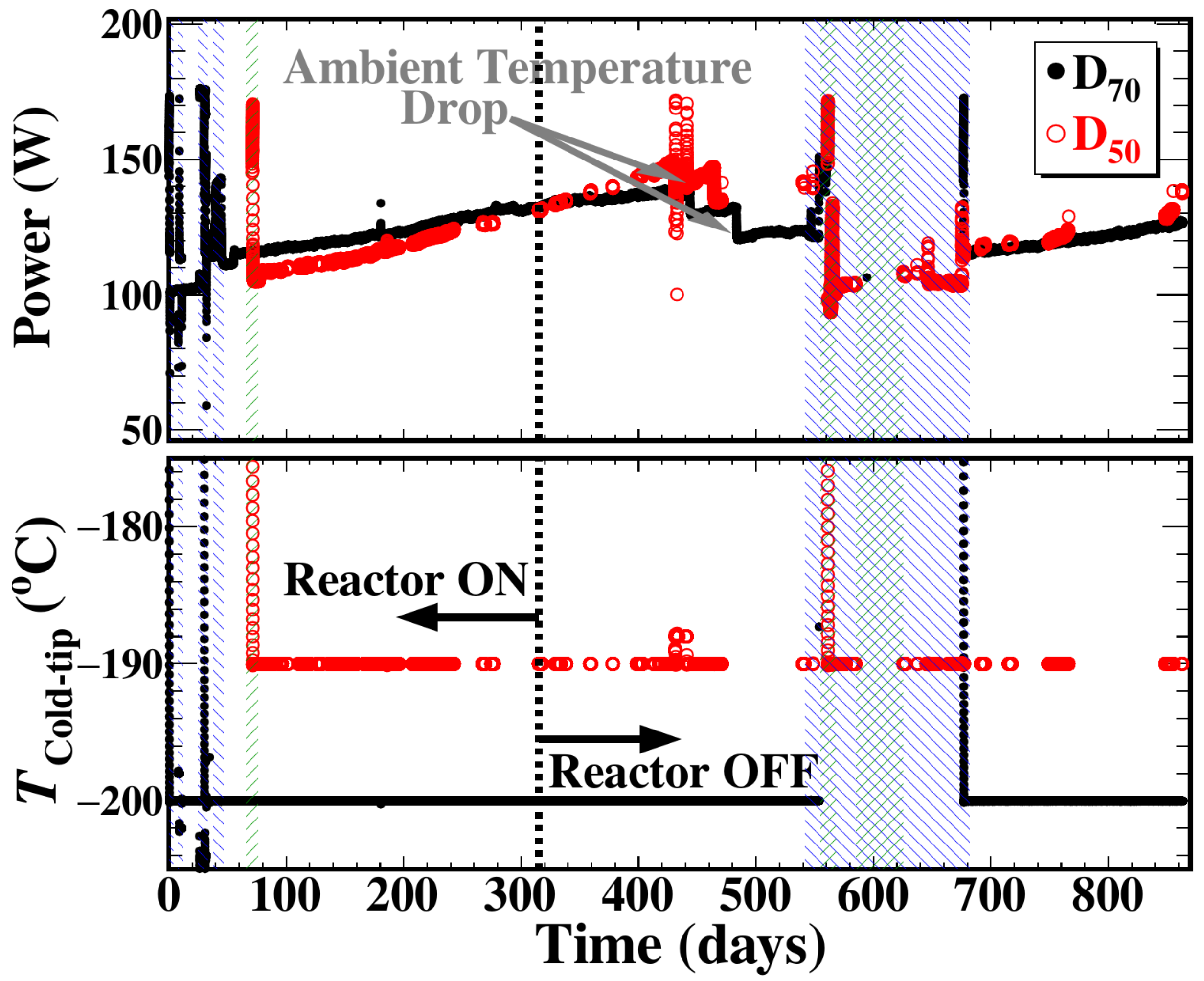} \\
  {\bf (b)}
  \includegraphics[width=8.2cm]{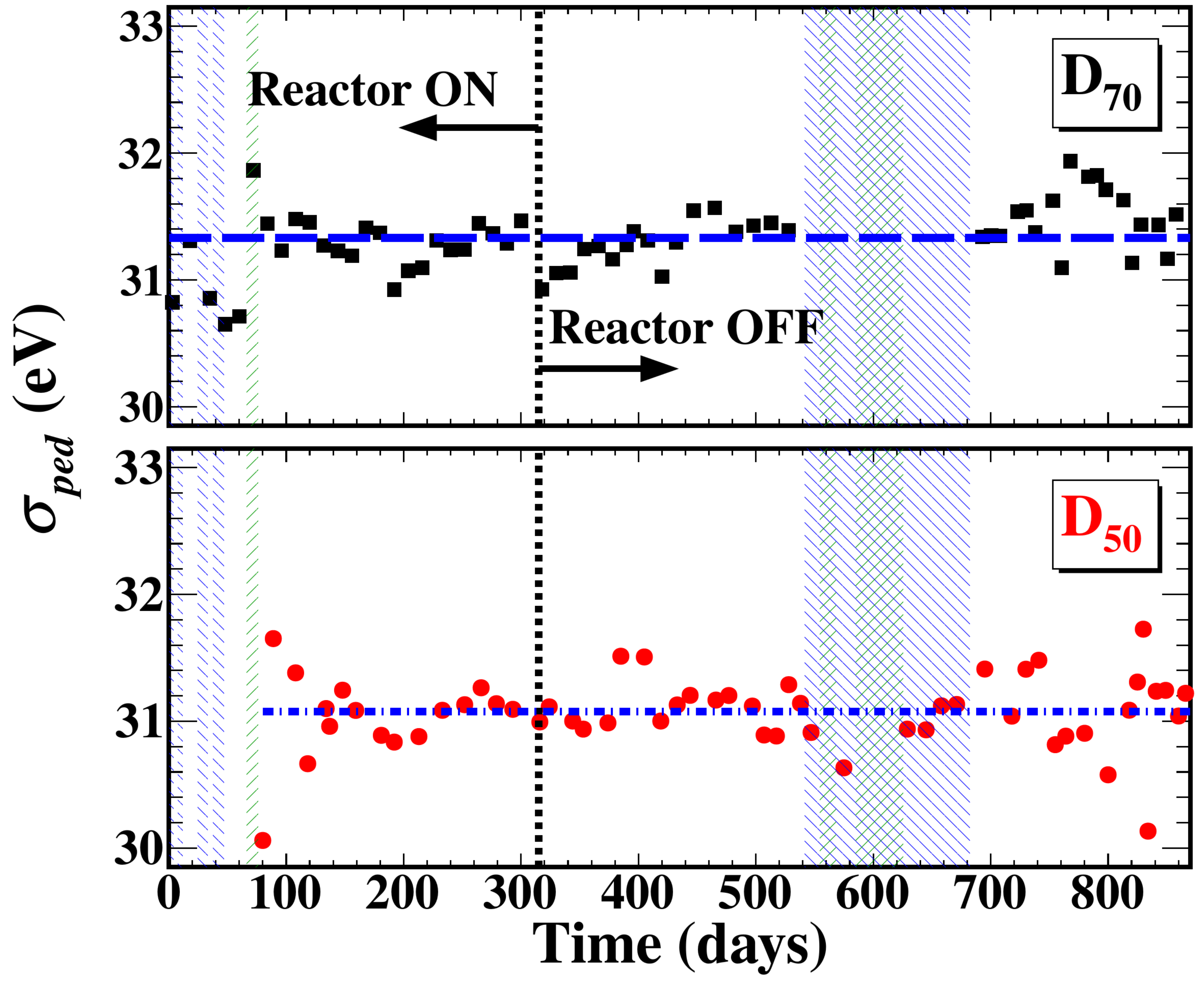}
  \end{center}
  \caption{(a) Time variations of the cold-tip
    temperature sensor located $\sim$ 54~cm from the
    Ge-crystal, and the power consumption for
    both $\BigDet$ and $\SmallDet$. All
    fluctuations are attributed to well-characterized
    hardware responses or ambient conditions.
    (b) The stability of the pedestal noise
    $\sigma_{ped}$ during the same DAQ periods. The
    measurements demonstrate that the inevitable
    variations in the cooler power have no
    measurable effect on pedestal stability.
    The shaded bands mark the periods during which
    data taking was temporarily suspended.
  }
  \label{fig::CP5}
\end{figure}

Independent of the DAQ system, the operating
conditions of the CP5+ cryocooler are
recorded for stability monitoring.
The measured parameters include
the temperatures of the controller,
compressor, cold head, and cold-tip,
together with the corresponding power
consumption. As an illustration, the
evolution of the cold-tip temperature
and power for the $\BigDet$ and $\SmallDet$
ECPCGe detectors during data taking
periods when both detectors are active
is shown in Figure~\ref{fig::CP5}(a).
The cold-tip is thermally coupled to
the PCGe diode and thus provides the most direct
measure of the crystal temperature.
Cooling from room temperature to the
operating set points (-200\textdegree C
for $\BigDet$ and -190\textdegree C for
$\SmallDet$) requires approximately
15~h, while warm-up takes about 12~h.
High voltage is applied to the PCGe detectors
only after thermal equilibrium is reached.
Proper compressor operation requires an
ambient temperature of $\sim$10-15\textdegree C,
maintained by an air-conditioning system
inside the shielding structure. The
observed power fluctuations in
Figure~\ref{fig::CP5}(a) are due to known
and well-understood ambient conditions, while
the cold-tip temperatures remain stable
throughout. The shaded bands denote
periods of suspended data taking due to
various reasons. 


Measurements of reactor $\nuAel$
requires long-duration data taking and
accurate comparison of data from 
the reactor ON and OFF periods. 
System stability, in
particular the detector threshold,
is of fundamental importance. The
pedestal noise levels are sampled at
0.1~Hz with the RT events. As an illustration,
the root-mean-square (RMS) of the pedestal of
the SA signals ($\sigma_{ped}$) at the highest
amplification (that is, the channel
which defines the threshold) for both
$\BigDet$ and $\SmallDet$ is displayed
in Figure~\ref{fig::CP5}(b). The data show
no correlation with the electro-cooler
power consumption and demonstrate
stable detector operation. The induced
fluctuations in the detector response
at a $\T0{=}$200~$\eVee$
are $\lesssim$~3.5~eV and $\lesssim$~2.8~eV
for $\BigDet$ and $\SmallDet$,
respectively, which are well below the
width of a single energy bin in the
subsequent analysis. Periods of
suspended or sub-optimal DAQ are shown
as shaded bands. These are consequences
of a variety of hardware
issues discussed in
Section~\ref{sect::raw}.


\begin{table*}
 \centering
 \caption{Summary of the key experimental
   configurations and performance parameters
   for the TEXONO ECPCGe $\nuAel$ experiment
   at KSNL.}
\begin{center}
\renewcommand{\arraystretch}{1.1}
\begin{tabular}{|l|c|c|}
\hline
 
Key elements                            & $\BigDet$                    & $\SmallDet$             \\ \hline \hline
                 
Reactor $\nuebar$-flux                  &\multicolumn{2}{c|}{$ 6.37 {\times} 10^{12} ~ \pcm2s1 $}   \\ \cline{2-3}

Total Sensor Mass (g)                   & 1434                         & 523                     \\
Fiducial Mass (g)                       & 1334 ($\pm$ 0.74\%)          & 472 ($\pm$ 1.04\%)      \\
Detector Threshold ($\eVee$)            & 200                          & 200                     \\
Resolution (RMS) at Threshold ($\eVee$) & 32.5                         & 32.3                    \\
Pedestal RMS Noise ($\eVee$)            & 31.3                         & 31.1                    \\
Test Pulser FWHM ($\eVee$)              & 70.2                         & 70                      \\
\multirow{3}{*}{DAQ Calendar Period}    & \multirow{3}{*}{Sept. 2020 - Apr. 2023}  & Sept. 2019 - Jun. 2020   \\
                                        &                              & Jan. 2021 - Apr. 2023   \\
                                        &                              & May-Aug. 2022           \\
DAQ Valid Data Reactor ON ($\kgd$)      & 242                          & 162                     \\ 
\hspace*{2.4cm} Reactor OFF ($\kgd$)    & 559.3                        & 254.4                   \\ \hline
\end{tabular}
\end{center}
\label{tab::ksdaq}
\end{table*}


\subsection{Data Analysis}
\label{sect::analys}

The data analysis procedures closely follow those
established in previous experiments with
Ge-detectors and configurations similar to that
shown in Figure~\ref{fig::expt}. Details can be
found in Refs.~\cite{TEXONO:2007xds,TEXONO:2016:AKSoma,
  TEXONO-PRL2025}. In this article, we present
measurements specific to the datasets on $\nuAel$
with $\BigDet$ and $\SmallDet$.

\subsubsection{Raw Data}
\label{sect::raw}

Data were collected at KSNL during calendar time
September 2020-April 2023 and January 2021-April
2023 from $\BigDet$ and $\SmallDet$, respectively.
Additional data were acquired with $\SmallDet$
operating in single-detector mode from September
2019 to June 2020, as well as during the period
May-August 2022 when only $\SmallDet$ was in
operation. The COVID-19 pandemic
lockdown posed severe restrictions
on the operation of the experiment. For extended
periods, filled hard disks could not be retrieved
and replaced, hardware malfunctions could not be
attended, and changing ambient conditions could
not be mitigated. These significantly reduced the
fraction of analyzable physics data to 242(559.3) and
162(254.4)~$\kgd$ under Reactor ON(OFF) conditions
for $\BigDet$ and $\SmallDet$,
respectively. The key characteristics and DAQ
information of both detectors are summarized
in Table~\ref{tab::ksdaq}.

\begin{figure}
  \begin{center}
    {\bf (a)}\\
    \includegraphics[width=8.2cm]{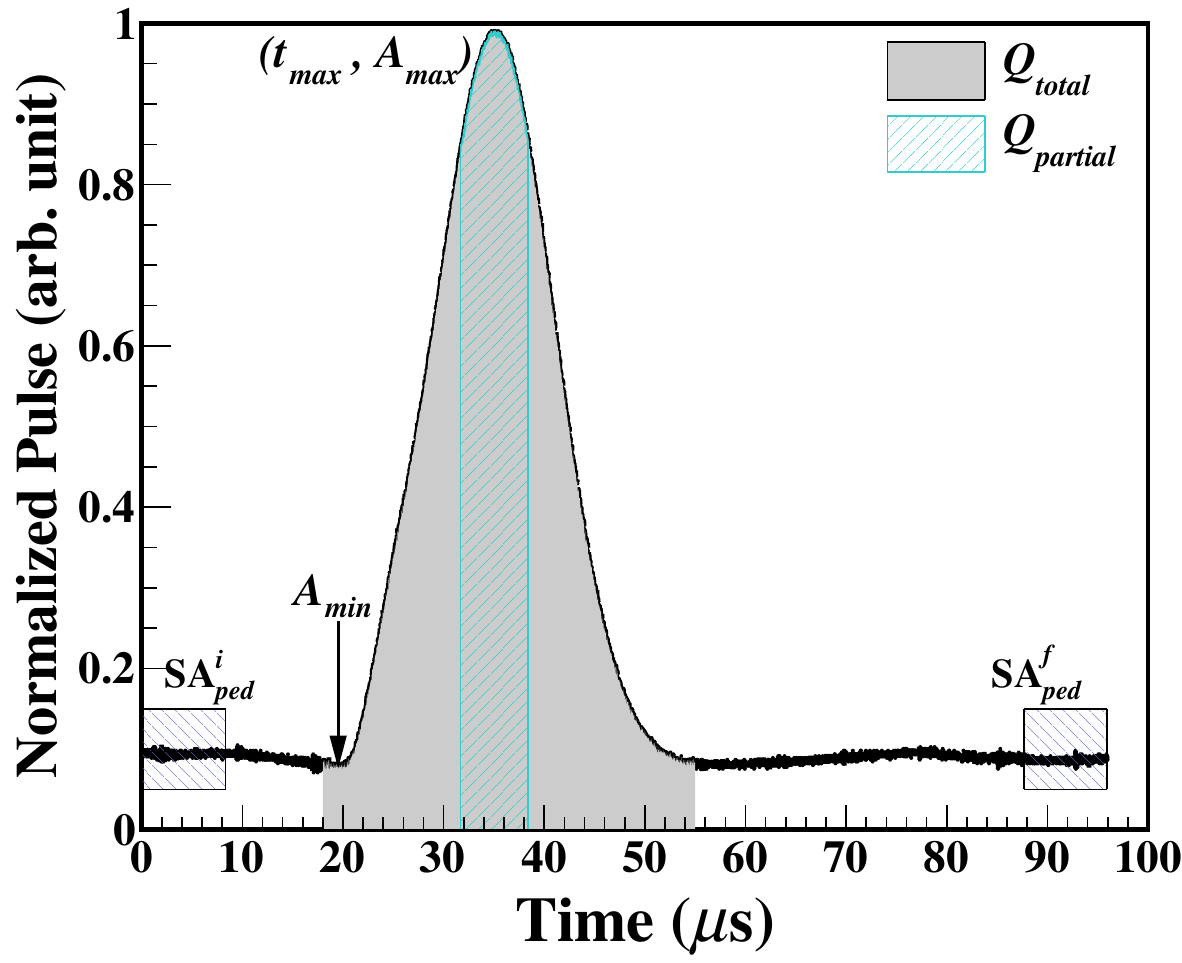} \\
    {\bf (b)}\\
    \includegraphics[width=8.2cm]{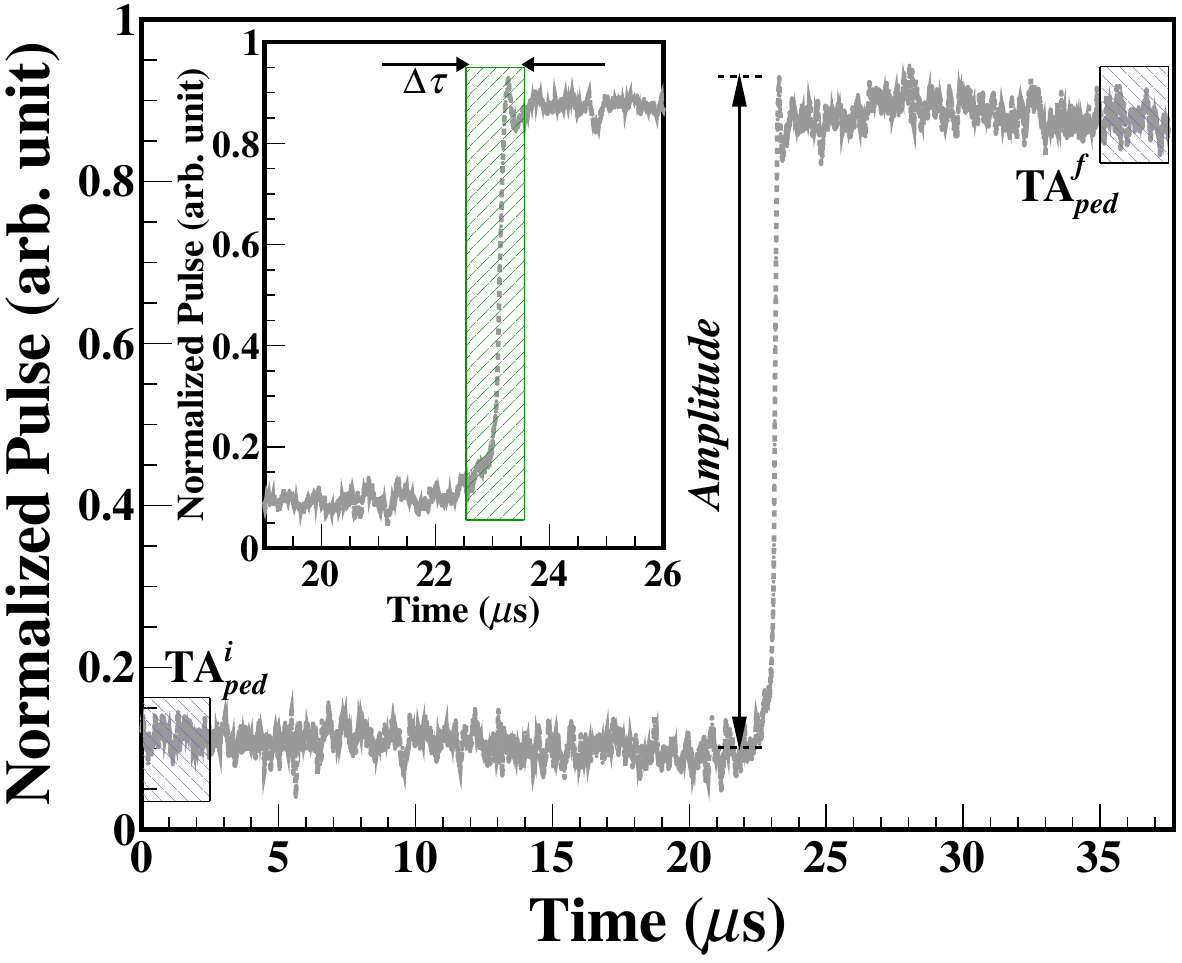} 
  \end{center}
  \caption{A typical (a) slow-pulse from the SA and
    (b) the corresponding fast-pulse from the TA for
    a 5~keV event, along with the pulse shape
    parameters that characterize the pulses. The
    inset in (b) illustrates the $\tau$
    behavior on an expanded time scale.}
  \label{fig::pulseshape}
\end{figure}

Typical SA (slow) and TA (fast) pulses recorded
by the DAQ system are depicted in
Figures~\ref{fig::pulseshape}(a) and \ref{fig::pulseshape}(b), respectively.
The relevant parameters to be derived for
subsequent analysis are also shown.


\subsubsection{Signal Selection}
\label{sect::signalselection}

Candidate signals due to neutrino interactions in
ECPCGe are uncorrelated with other detector components.
The raw signals undergo a sequence of selection
procedures, from which the candidate events are
identified.

The selection procedures, described in the following 
paragraphs, assign CR and AC tags to the
events, and distinguish between bulk
and surface ($B$ versus $S$) interactions. The events are categorized
as ``AC$^{-(+)}\otimes$CR$^{-(+)}\otimes$B$_{0}$/S$_{0}$'', where the
superscript $-(+)$ denotes anti-coincidence(coincidence)
with Ge signals~\cite{TEXONO:2016:AKSoma}. Candidate
events from neutrino- or DM-induced interactions
in the bulk region of the ECPCGe are uncorrelated with
other detectors and are therefore extracted from the
$\VVB$ category, where B$_{0}$ represents ``raw-bulk''
events.

\begin{figure}
  \begin{center}
  {\bf (a)} \\
  \includegraphics[width=8.2cm]{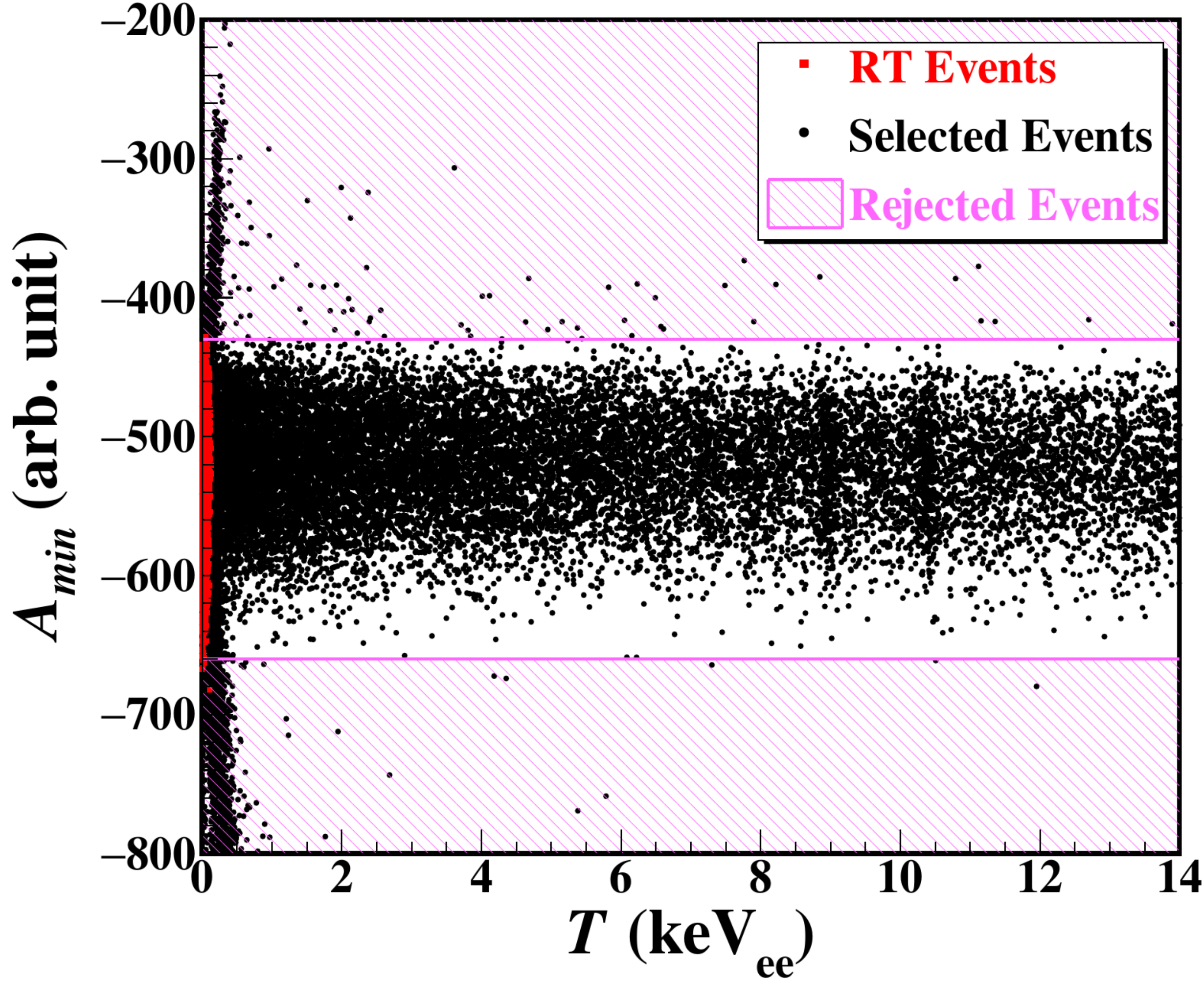} \\
  {\bf (b)} \\
  \includegraphics[width=8.2cm]{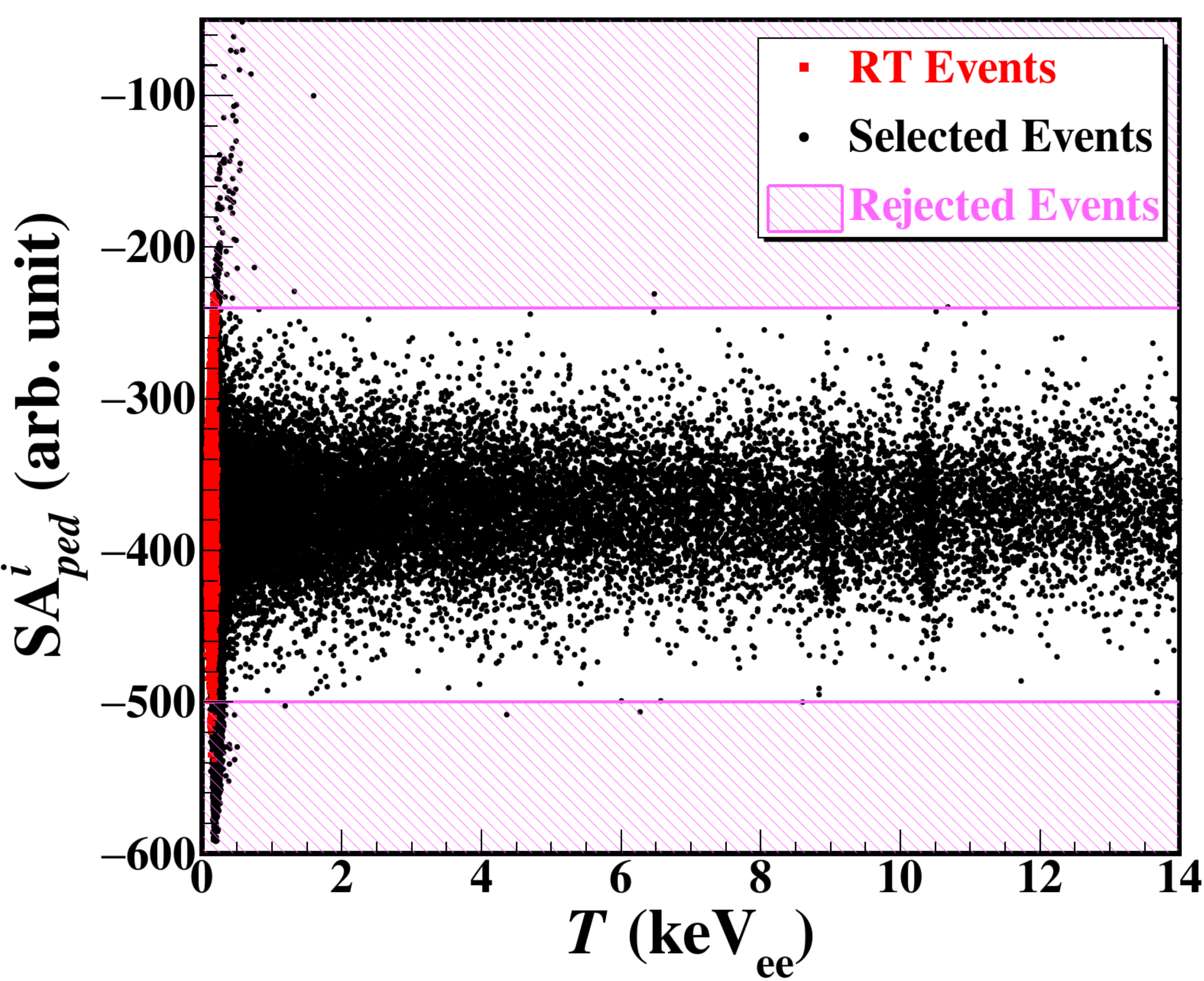} \\
  {\bf (c)} \\
  \includegraphics[width=8.2cm]{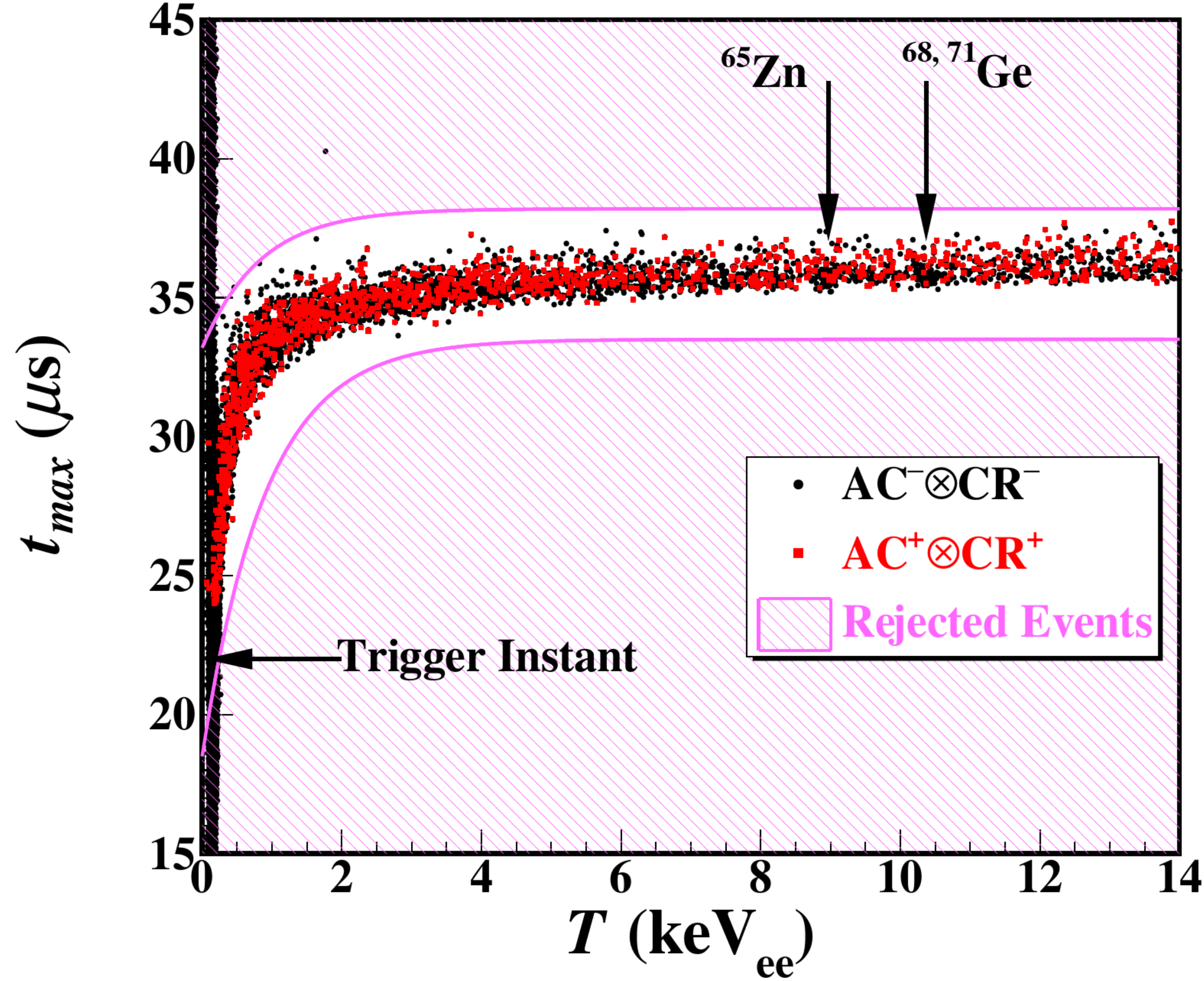} 
  \end{center}
  \caption{Typical distributions of (a) $A_{min}$,
    (b) SA$_{ped}^{i}$, and (c) $t_{max}$ versus
    measured energy $T$, following the notations of
    Figure~\ref{fig::pulseshape}. Events in the
    shaded regions originate from pedestal noise
    and fluctuations are discarded. The red
    data points are for signal efficiencies
    measurements, with RT events in (a) and (b),
    and with $\TT$ events in (c). The structures
    in the $t_{max}$ versus $T$ distribution in (c)
    are due to the SA shaping time at 6~$\mu s$.}
  \label{fig::basic}
\end{figure}

Signal inefficiencies due to DAQ and CR systems are independent
of the Ge-signatures. These inefficiencies are measured with survival
fractions of RT events. Triple coincidence events
in the $\TT$ category, however, are samples of pure physics events.
They are used to evaluate the signal efficiencies of Ge pulse
shape related cuts, like in ``noise-edge'' (NE) selection.

\begin{enumerate}
\item {\it Baseline selection}:
  Valid events would have pedestal levels within
  a range of $\approx\pm$4$\sigma$, as illustrated in
  Figures~\ref{fig::basic}(a) and~\ref{fig::basic}(b).
  Spurious noise as well as pile-up events are effectively
  filtered out. The maxima of the SA pulse characterized
  by the amplitude ($A_{max}$) and the time ($t_{max}$),
  would follow a definite
  pattern relative to the trigger instant. The selection
  is depicted in Figure~\ref{fig::basic}(c).
  
\item {\it Cosmic-ray veto}:
  Cosmic-ray-induced events are identified by
  the CR system with plastic scintillator panels,
  which have timing response of 10-100~ns, different
  from the ECPCGe SA trigger ($\sim$6~$\mu$s).
  The time difference $\Delta t$ between the
  trigger and the closest CR events versus $T$ is
  shown in Figure~\ref{fig::cr+bs}(a),
  revealing the time structure of the coincidence
  band. RT events are used to quantify the
  wrong-rejection efficiency of $\VVB$ samples,
  yielding a signal acceptance efficiency of
  $\sim$92\%. The CR detection efficiency of
  $\sim$93-95\% is derived from AC$^{+}$ samples
  with large ($\approx$20~$\MeVee$) energy
  deposition in the AC detectors.
  
\item {\it Anti-Compton veto}:
  The NaI(Tl) AC detector has an
  analysis threshold of $\sim$5~$\keVee$. The $\VVB$
  signal efficiency, measured by RT events, is
  close to unity. The $\nuAel$ events are incorrectly
  rejected at $\sim$0.5\%.
  
\item {\it Bulk-Surface events selection}:
  The Li-diffused $n^{+}$ surface ($S$) layer of a
  $p$-type PCGe detector has a thickness of
  $\sim$1~mm ($\pm$ 10\%)~\cite{Jiang::CPC2016,MA2017:CDEX,AGUAYO::NIMA2013176},
  consisting of a dead layer and a region with
  suppressed and slower charge collection. Adopting
  the previously developed algorithms of
  Refs.~\cite{TEXONO:2013bju,Yang:2016crf,Wang:2024phr},
  the $B$ and $S$ events are differentiated with the
  $\tau$ of the TA pulses, as illustrated
  in Figure~\ref{fig::cr+bs}(b). Events shown in red
  correspond to those from a test pulser programmed
  to have an identical shape to $B$ events,
  serving as a reference for signal efficiency
  measurements. The $B/S$ events are indistinguishable
  in the SA pulses, which are shaped by a much slower
  time constant of 6~$\mu$s. The selection shown in 
  Figure~\ref{fig::cr+bs}(b) produces
  samples of B$_{0}$ events,
  which are contaminated near-threshold
  energies by leakage from surface background events.
  Correction procedures are
  devised~\cite{TEXONO:2013bju,Yang:2016crf,Wang:2024phr}
  to obtain ``corrected-bulk'' (B$_{\rm c}$)
  samples. 

  \begin{figure}
  \begin{center}
  {\bf (a)} \\
  \includegraphics[width=8.2cm]{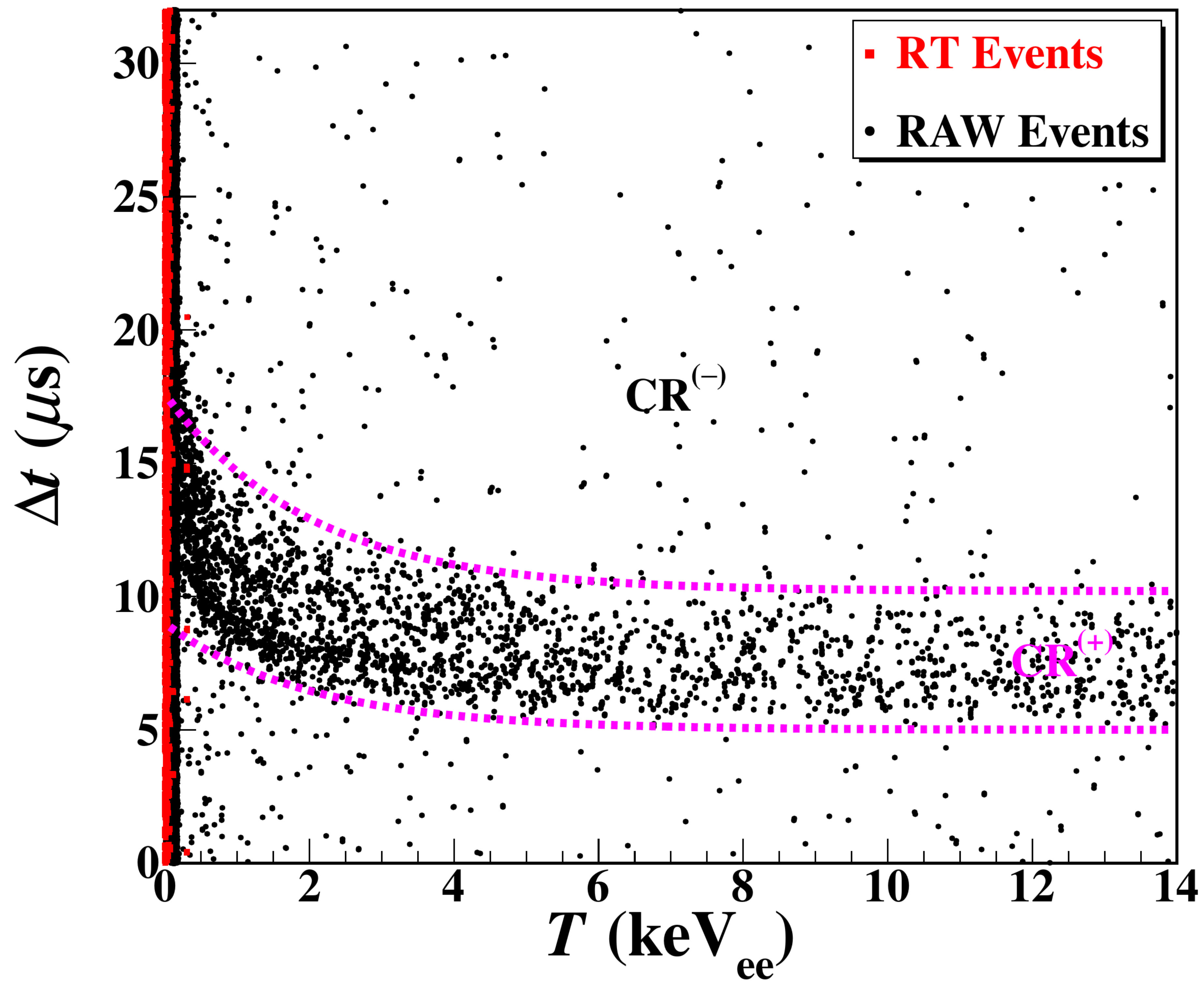} \\
  {\bf (b)} \\
  \includegraphics[width=8.2cm]{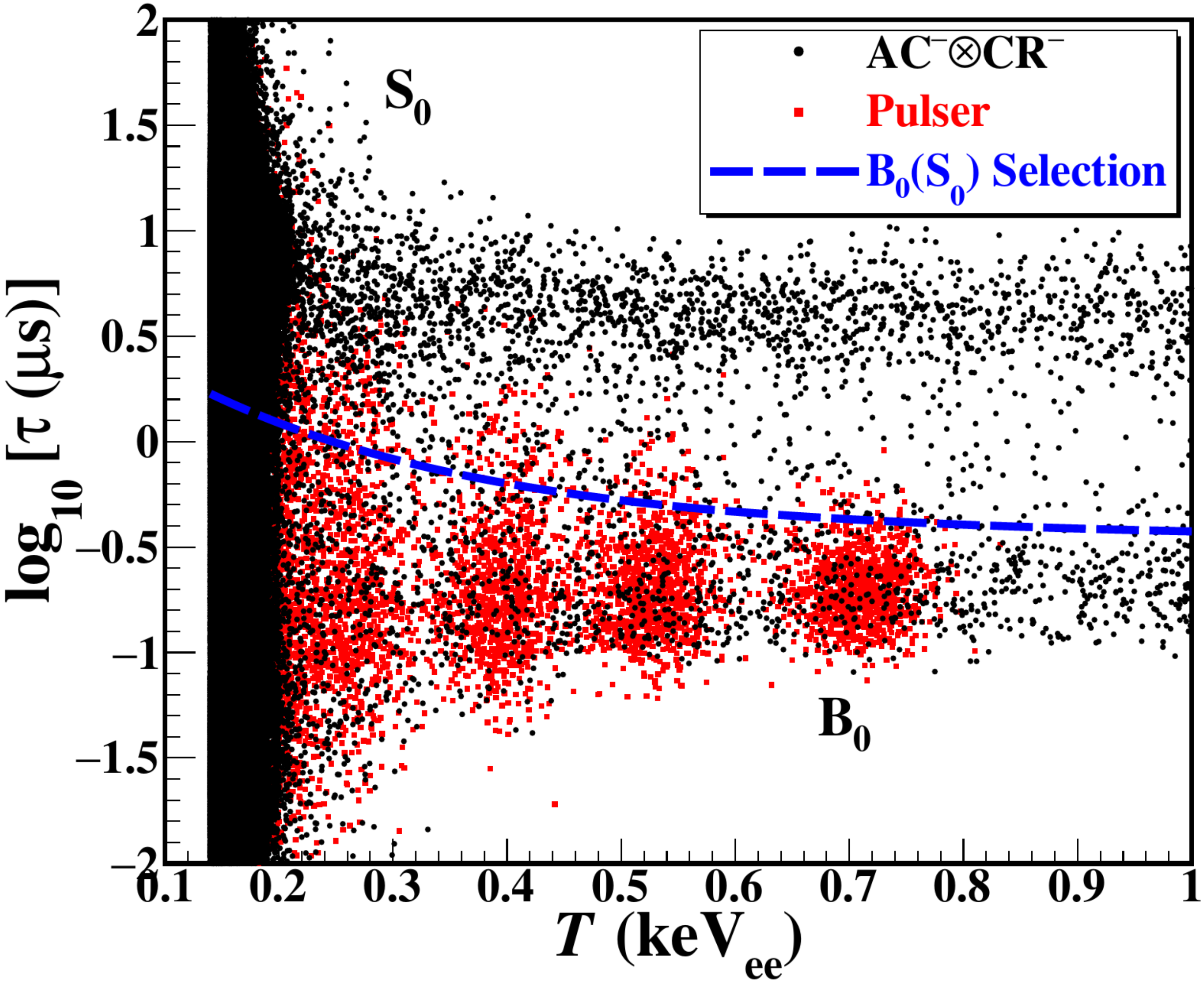} 
  \end{center}
  \caption{(a) Typical time differences ($\Delta t$) versus
    $T$ between CR pulses and the DAQ trigger
    instants produced by the slow SA signals. Samples
    within the dotted magenta contours are identified
    as CR$^{+}$ events and are rejected. (b) The $\tau$
    distributions for physics $\VV$ events. The
    B$_0$(S$_0$) selection contour is indicated by the
    dashed blue line. The red data points are for signal
    efficiency measurements, from RT events in (a)
    and test pulser events in (b).}
  \label{fig::cr+bs}
  \end{figure}
  
\item {\it Noise-edge selection}:
  The discriminator level for the DAQ trigger is
  set well below the detector NE,
  defined as the energy below which electronic
  noise dominates the event rate over
  physics events. These sub-NE events are saved
  for possible future research on differentiation
  of physics versus noise events via advanced
  pulse shape analysis techniques. The
  effective trigger threshold is at about
  150~$\eVee$ resulting in a DAQ
  rate of about 50~Hz. 
The NE selection, as depicted in Figure~\ref{fig::ne}, 
is applied to decouple the sub-NE 
pedestal noise ``self-trigger'' and other
  spurious events from the $\VVB$ candidate signal samples. 
The well understood $\TT$ background samples are used for the 
measurement of the signal efficiency. 
The resulting detector threshold for physics analysis is $\T0 {=} 200 ~ \eVee$.
\end{enumerate}

\begin{figure}
  \begin{center}
    \includegraphics[width=8.2cm]{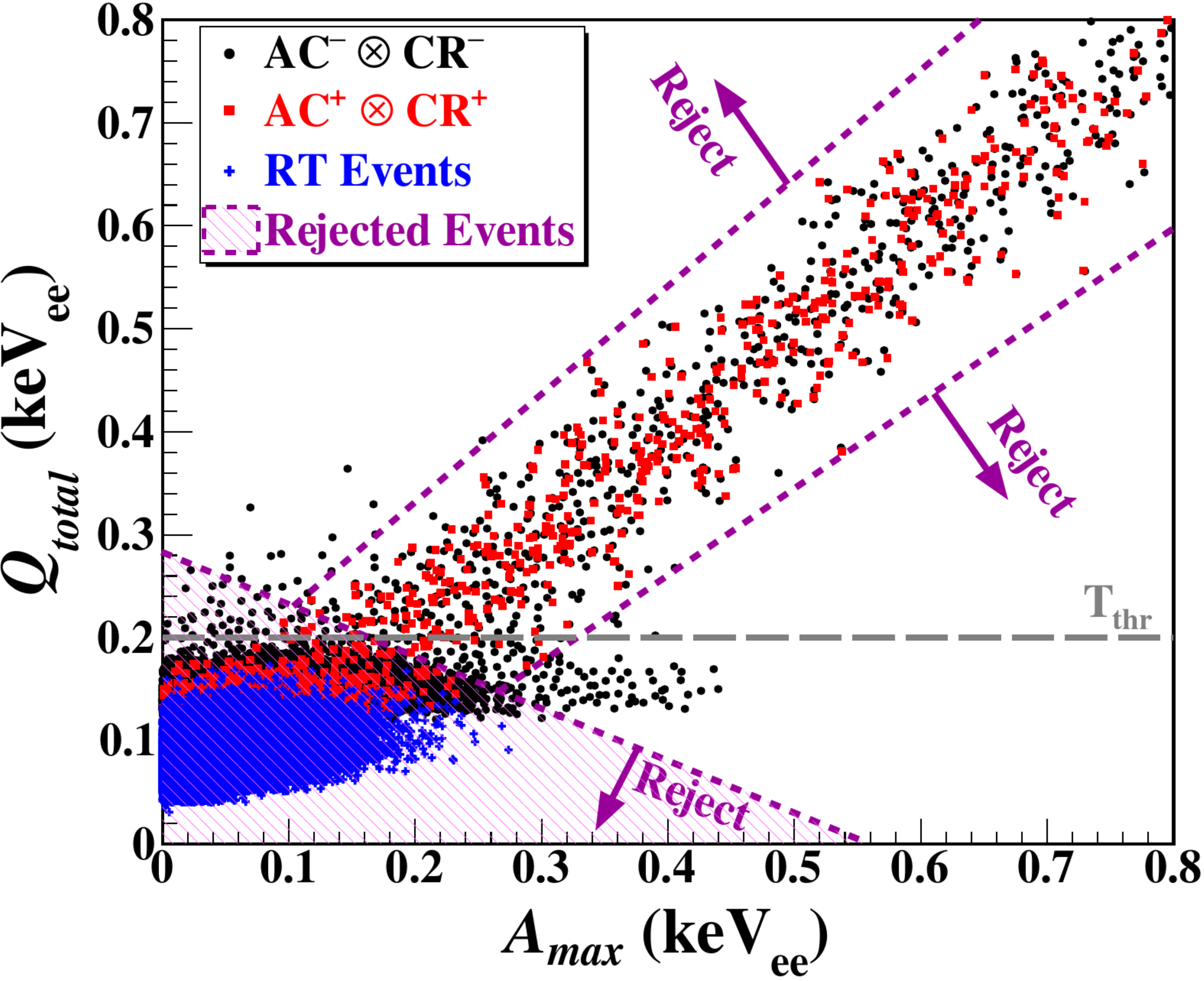} 
  \end{center}
  \caption{Scatter plot of $Q_{total}$ versus $A_{max}$
    for a representative $\VV$ sample, from which
    the threshold cut is
    defined. The red and blue data points correspond to
    $\TT$ and RT events, respectively, and are used for
    signal efficiency measurements. }
  \label{fig::ne}
\end{figure}


\section{Results}
\label{sect::results}

\subsection{Reactor Spectra}
\label{sect::reac::spec}

Nuclear power reactors are intense and
steady sources of $\nuebar$, making them
ideal for low energy neutrino studies.
The KSNL is located at 28~m from Core-1 of the
Kuo-Sheng Nuclear Power Station, which
operates at a nominal thermal power of
2.9~GW~\cite{TEXONO:2018}. The $\nuebar$
spectra is derived following the models
of Figure~5 of Ref.~\cite{TEXONO:2007xds}. 
The best estimate of the anomalous
``5-MeV bump'' is
included~\cite{DayBay::PRL2025}. The total
$\nuebar$-flux is
6.37$\times10^{12}$~$\pcm2s1$.

\begin{figure}
  \begin{center}
    \includegraphics[width=8.2cm]{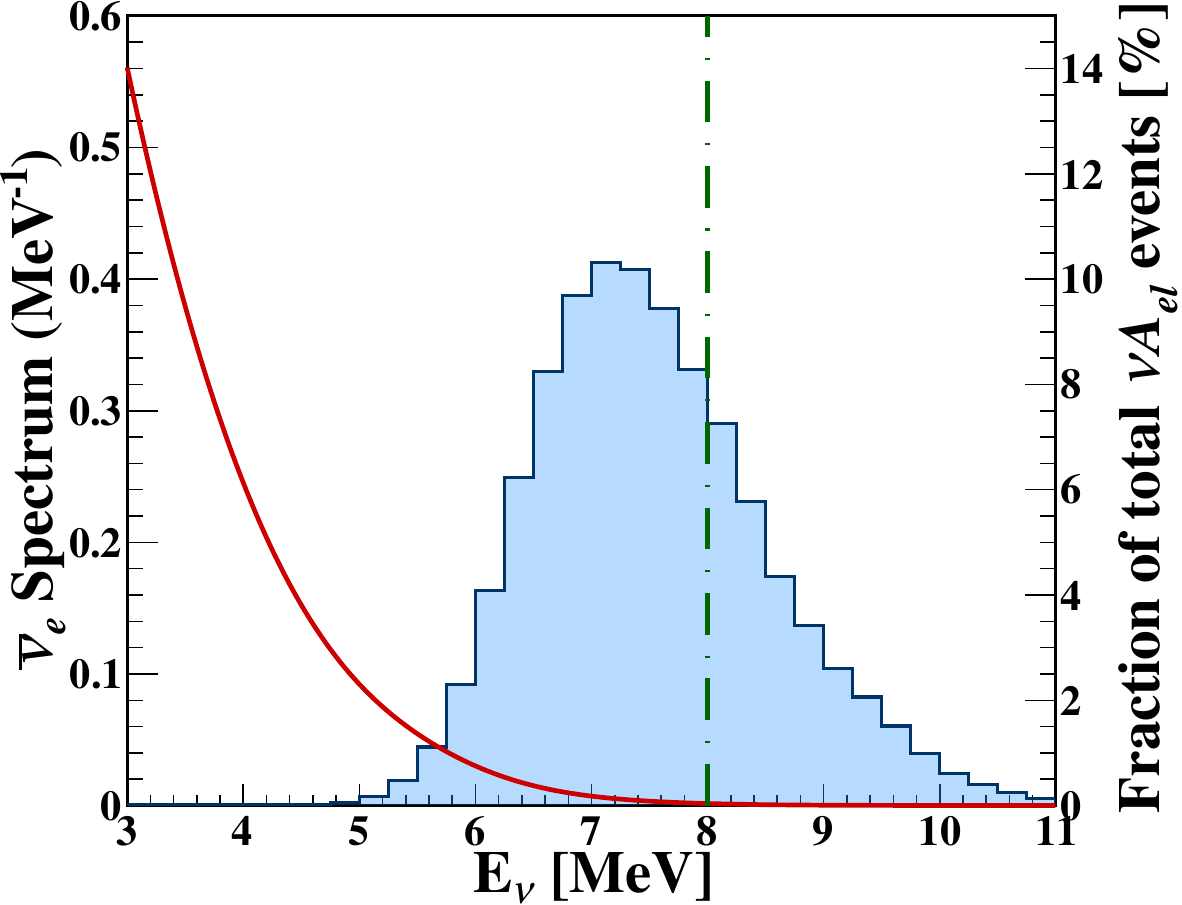} 
  \end{center}
  \caption{Reactor $\nuebar$ spectrum (red) at
    energies above 3~MeV under typical operation.
    The blue histogram depicts the distribution
    of $\Enu$, which gives rise to the expected
    SM $\nuAel$ events at $\T0 {=} 200~\eVee$ in ECPCGe.
    About 29.4\% of the events are produced by
    $\nuebar$ with $\Enu>$8~MeV.}
  \label{fig::nuspect}
\end{figure}

Reactor $\nuebar$ analyses are traditionally
restricted to energies of $\Enu<$8~MeV.
Recent measurements by the Daya
Bay~\cite{DayaBay::PRL2016,DayaBay::PRL2019,PRL:2022:DayB}
and JUNO~\cite{JUNO::PRNP2022} experiments
extend to the 8-11~MeV range. The total
cross section for $\nuAel$ is proportional
to $\Enu^{2}$ and the maximal $\TNR$ is
constrained by Eq.~(\ref{max::reco}).
Consequently, the high energy $\nuebar$-spectra
provide non-negligible 
contributions, particularly when only the tails
of the recoil energy are measured. 
The high energy tail of the $\nuebar$-spectrum at KSNL
and the portion which interacts via
$\nuAel$ to provide nuclear recoil
events at $\T0{=}$200~$\eVee$ are
depicted in Figure~\ref{fig::nuspect}.
The $\nuebar$-spectra in
the 8-11 MeV range and from the 5-MeV bump
contribute about 29.4\% and 2.25\%,
respectively, of the total $\nuAel$ events.


\subsection{Quenching Factor}
\label{sect::qf}

The ECPCGe measures ionization in the form of
electron-hole pairs, while the final-state
observable in $\nuAel$ arises from $\TNR$ of the nucleus. 
The equivalent electron-recoil yield is characterized
by the QF. Accurate knowledge
of the QF is essential for interpreting $\nuAel$
data, particularly in future precision
measurements.

\begin{figure}
  \begin{center}
    \includegraphics[width=8.2cm]{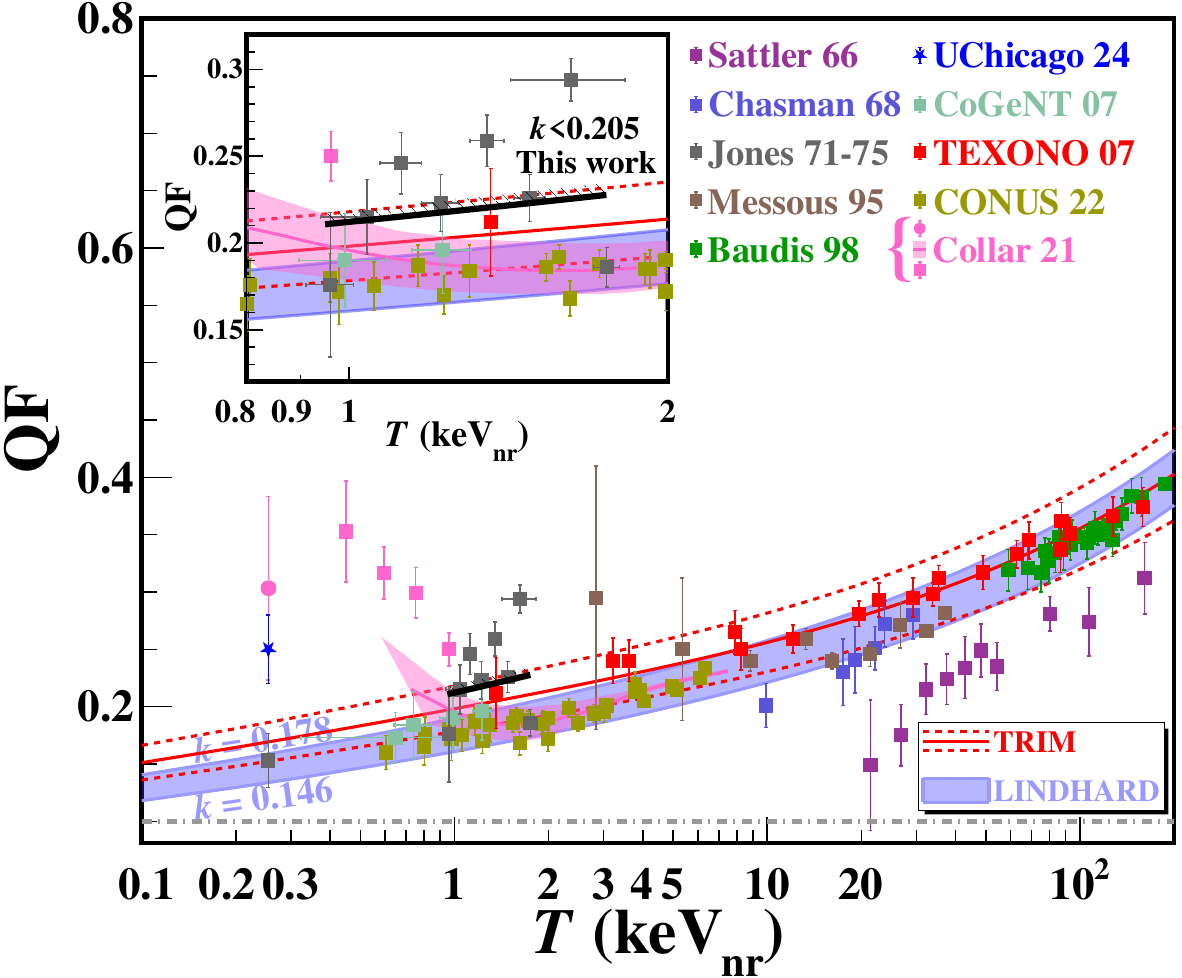} 
  \end{center}
  \caption{Summary of 
QF data~\cite{TEXONO:2016:AKSoma,JICollar_QF_PRD21,
Bonhomme:conus,PRD::2024::KARLJI}
    in Ge ionization detectors. Superimposed are
    values derived with the TRIM
    software~\cite{Ziegler1998:TRIM} (red curve with a
    $\pm$10\% uncertainty shown by dotted lines) and
    the Lindhard model~\cite{LinHD:19621,LinHD:19622,LinHD:1964}
    at a commonly adopted parameterizations
    $k$($\equiv$0.162), including an uncertainty of
    $\pm$10\% (light blue band). There is good
    agreement between the data and model predictions
    across the relevant $\TNR$. The
    outlying data points are from Ref.~\cite{JICollar_QF_PRD21}, 
obtained with three distinct measurement methods (denoted by the
    light pink band and data associated
    with ``Collar 21''). These results have been
    tested and rejected by subsequent
    measurements~\cite{CONUS-2,Bonhomme:conus,LiLong::2022,
      TEXONO-PRL2025}. The upper limit at 90\% CL
    corresponding to $k<$0.205 derived from this
    work, is depicted in the inset.}
  \label{fig::qf}
\end{figure}

A compilation of existing QF measurements in Germanium
at liquid nitrogen temperature~\cite{TEXONO:2016:AKSoma},
including recent updates~\cite{Bonhomme:conus,JICollar_QF_PRD21},
is presented in Figure~\ref{fig::qf}. Superimposed are
the theoretical predictions from simulations using the
TRIM package~\cite{Ziegler1998:TRIM} and from the
semi-empirical Lindhard
model~\cite{LinHD:19621,LinHD:19622,LinHD:1964}.


The TRIM-based QF prediction is presented
as the red solid contour, with a conservative
systematic uncertainty of $\pm$10\% is
indicated by the dotted lines. The TRIM simulations become
increasingly model-dependent and poorly constrained
with $\TNR$ at sub-keV~\cite{JLIAO::PRD2021:mgdl}.
The Lindhard model is presented with the benchmark
choice of $k$$\equiv$0.162~\cite{Bonhomme:conus,CONUS-2,TEXONO-PRL2025},
which is adopted in this analysis. The blue band 
corresponds to an uncertainty of $\pm$10\% in $k$.

A recent measurement~\cite{PRD::2024::KARLJI} of QF{=}0.25$\pm$0.03
at 254~eV$_{\rm nr}$ lies below the energy range relevant 
to this analysis, but it reveals the necessity of further
research to characterize the low energy responses in Ge.

\subsection{Signal Efficiencies}
\label{sect::sig::eff}

The $T$-independent contributions to
$\VVB$ signal efficiencies are measured from the
survival probabilities of RT events and are
summarized in Table~\ref{tab::sigeff}. The

\begin{figure}
  \begin{center}
    {\bf (a)}\\
    \includegraphics[width=8.2cm]{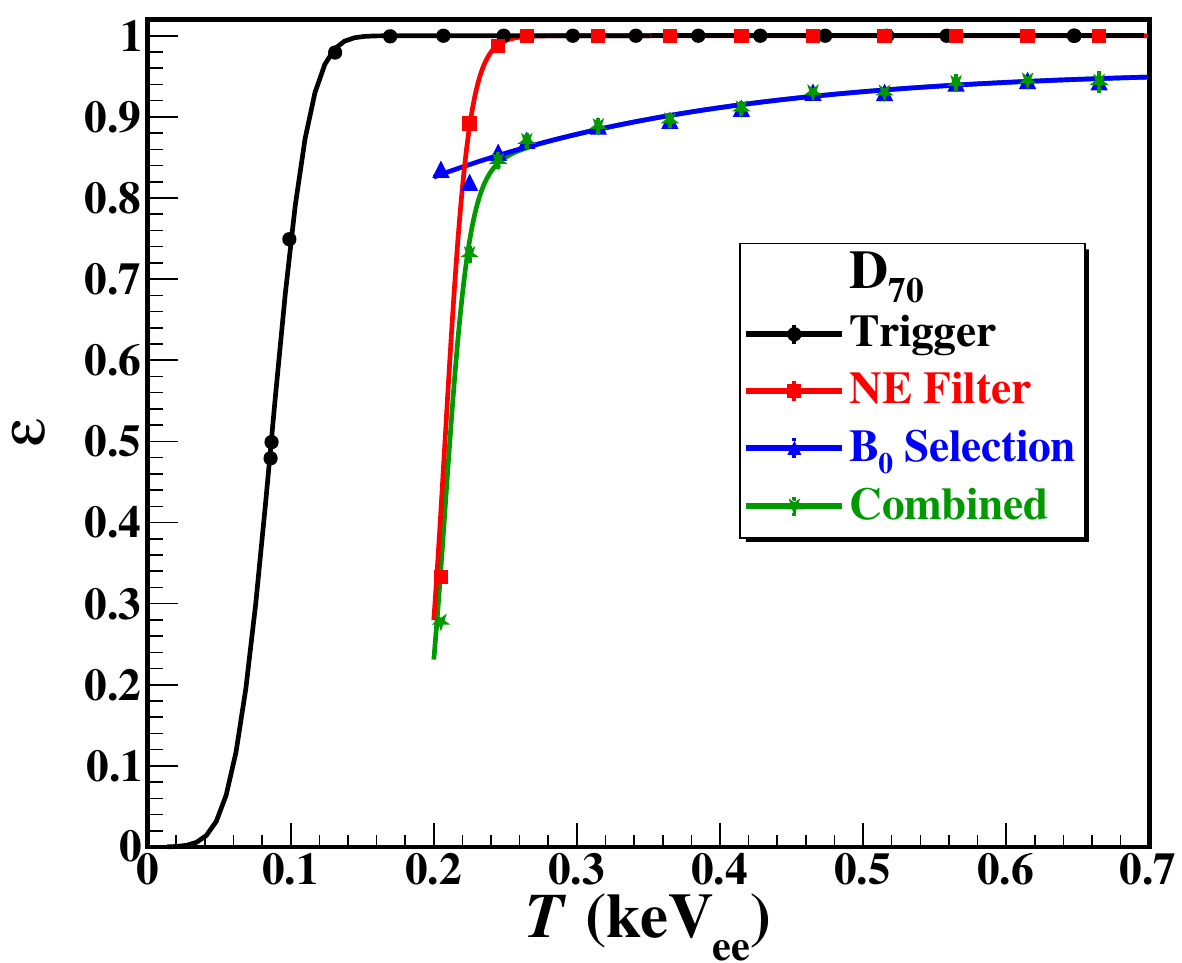} \\
    {\bf (b)}\\
    \includegraphics[width=8.2cm]{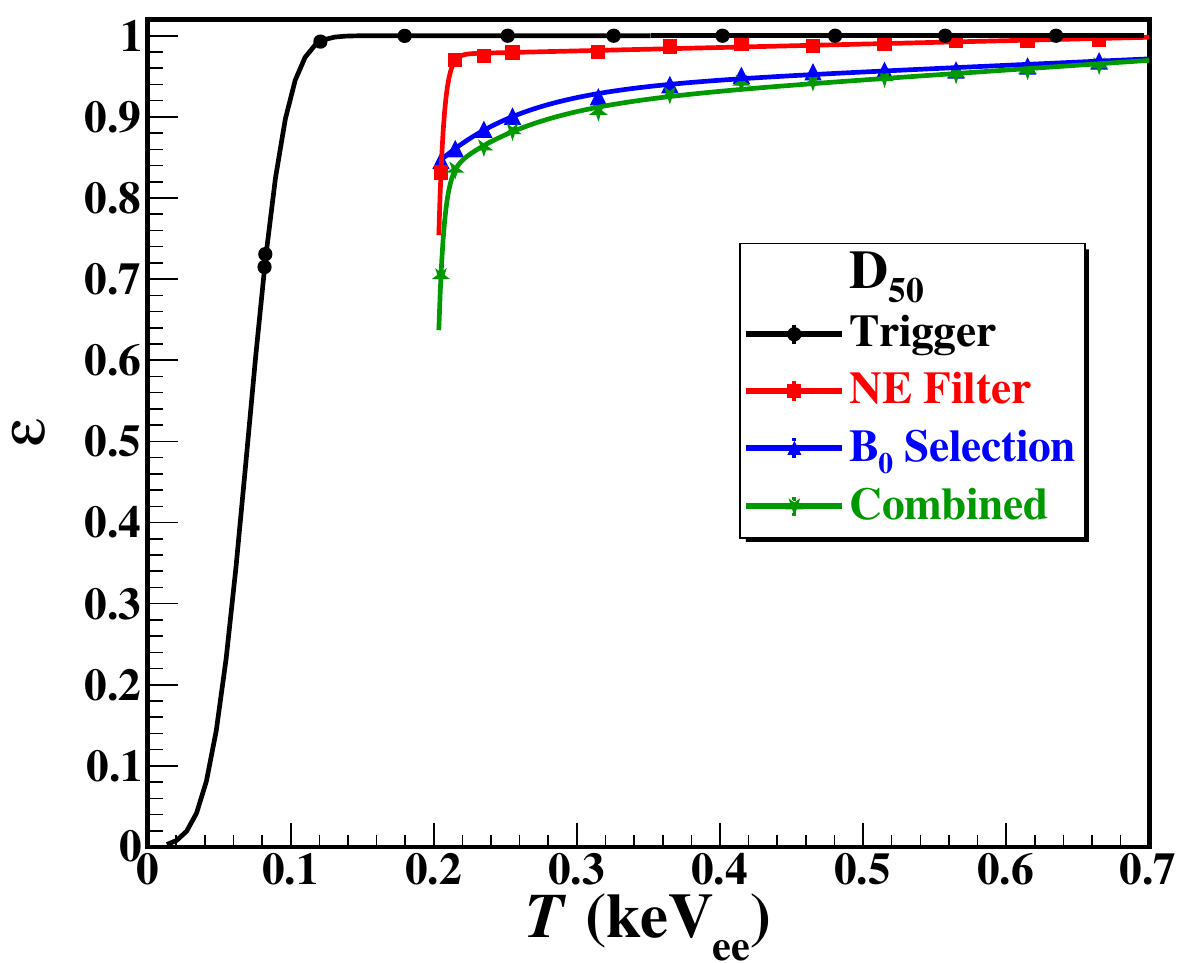} 
  \end{center}
  \caption{The $T$-dependent signal efficiency
    factors in the sub-keV energy region for (a)
    $\BigDet$ and (b) $\SmallDet$. The values at
    200~$\eVee$, together with the $T$-independent
    efficiencies, are tabulated in
    Table~\ref{tab::sigeff}.}
  \label{fig::sigeff}
\end{figure}

The $T$-dependent channels are displayed in
Figures~\ref{fig::sigeff}(a) and \ref{fig::sigeff}(b) for $\BigDet$
and $\SmallDet$, respectively. The discriminator
level for the DAQ trigger is set sufficiently
low to record 150~$\eVee$ events at full
efficiency to allow sampling of electronic
noise events for possible future R\&D. The
efficiency is derived from test pulser
events modeled from the SA pulse shapes.
Efficiencies due to the $B/S$
selection shown in Figure~\ref{fig::cr+bs}(b) are also measured
with test pulser events modeled to TA pulses.
The NE selection of Figure~\ref{fig::ne}
defines the physics analysis threshold. 
The efficiencies are derived from {\it in situ} $\TT$ events.
The values at $\T0{=}$200~$\eVee$ for both detectors are summarized
in Table~\ref{tab::sigeff}.

\subsection{Systematic Uncertainties}
\label{sect::sys::uncer}

The analysis results to be discussed in the following
sections indicate that statistical uncertainties dominate
the sensitivity of this measurement. The leading
contributions to the systematic uncertainties arise
from the evaluation of $\phi$($\nuebar$).
 An uncertainty of 5\% is assigned to the spectra
at $\Enu {<} 8~{\rm MeV}$.
This is comparable to, and more conservative than,
  those used in similar experiments~\cite{DRESDEN-1,
    Hakenmuller:2019ecb,CONNIE:2019xid,PRD::2015FMGE}.
  For the higher-energy region of 8-11~MeV, a larger
  uncertainty of 30\% is taken following
  Refs.~\cite{DayaBay::PRL2016,DayaBay::PRL2019,
    PRL:2022:DayB,JUNO::PRNP2022}, while a 40\%
  uncertainty is estimated for the 5-MeV bump region
  based on Ref.~\cite{DayBay::PRL2025}.
These contributions give rise to an uncertainty of 9.5\% 
to the expected SM $\nuAel$ rate at  $\T0 {=} 200 ~ \eVee$.

The systematic uncertainties associated with the signal
efficiencies are summarized in Table~\ref{tab::sigeff}.
The leading contribution arises from the $B/S$ selection,
amounting to 0.79\% and 0.21\% for $\BigDet$ and
$\SmallDet$, respectively, at the $\T0{=}$200~$\eVee$.


\begin{table}
 \centering
 \caption{Summary of the signal selection
   efficiencies and systematic uncertainties
   for $\BigDet$ and $\SmallDet$. The
   $T$-dependent selections at $\T0{=}$200~$\eVee$
   are listed. The complete
   $T$-dependence can be read from
   Figure~\ref{fig::sigeff}.}
\begin{center}
\renewcommand{\arraystretch}{1.1}
\begin{tabular}{|l|c|c|}
\hline
\multicolumn{3}{|c|}{ {\bf Detector:  $\BigDet$ } }\\ \hline
Signal Selection                  & ~~ $\varepsilon$ (\%) ~~ & Uncertainty (\%) \\ \hline 
\multicolumn{3}{|c|}{$T$-Independent} \\ \hline
~~ Data Acquisition               & 99.77            & $<$ 0.1 \\
~~ Basic Data Quality             & 99.64            & $<$ 0.1 \\
~~ Cosmic-Ray Veto $\CRV$         & 88.57            & $<$ 0.1 \\
~~ Anti-Compton Veto $\ACV$       & 99.84            & $<$ 0.1 \\
\multicolumn{1}{|r|}{Combined}    & 87.91            & $<$ 0.1 \\ \hline

\multicolumn{3}{|c|}{$T$-Dependent (at 200~$\eVee$)} \\ \hline
~~ Trigger                       & 99.99            & $<$ 0.1 \\
~~ Pedestal Noise Filter         & 33.25            & $<$ 0.1 \\ 
~~ Bulk/Surface Event Selection  & 83.56            & 0.79     \\
\multicolumn{1}{|r|}{Combined}   & 27.78            & 0.95    \\ \hline \hline

\multicolumn{3}{|c|}{ {\bf Detector:  $\SmallDet$ } }\\ \hline
Signal Selection                  & ~~ $\varepsilon$ (\%) ~~ & Uncertainty (\%) \\ \hline 
\multicolumn{3}{|c|}{$T$-Independent} \\ \hline
~~ Data Acquisition               & 99.77            & $<$ 0.1 \\
~~ Basic Data Quality             & 99.67            & $<$ 0.1 \\
~~ Cosmic-Ray Veto $\CRV$         & 92.30            & $<$ 0.1 \\
~~ Anti-Compton Veto $\ACV$       & 99.94            & $<$ 0.1 \\

\multicolumn{1}{|r|}{Combined}    & 91.73            & $<$ 0.1 \\ \hline

\multicolumn{3}{|c|}{$T$-Dependent (at 200~$\eVee$)} \\ \hline
~~ Trigger                       & 99.99            & $<$ 0.1 \\
~~ Pedestal Noise Filter         & 83.15            & $<$ 0.1 \\ 
~~ Bulk/Surface Event Selection  & 84.63            & 0.21    \\
\multicolumn{1}{|r|}{Combined}   & 70.36            & 0.25    \\ \hline

\end{tabular}
\end{center}
\label{tab::sigeff}
\end{table}


\subsection{Measured Spectra}
\label{sect::mes::spec}

\subsubsection{Background rejection}

\begin{figure}
  \begin{center}
    \includegraphics[width=8.2cm]{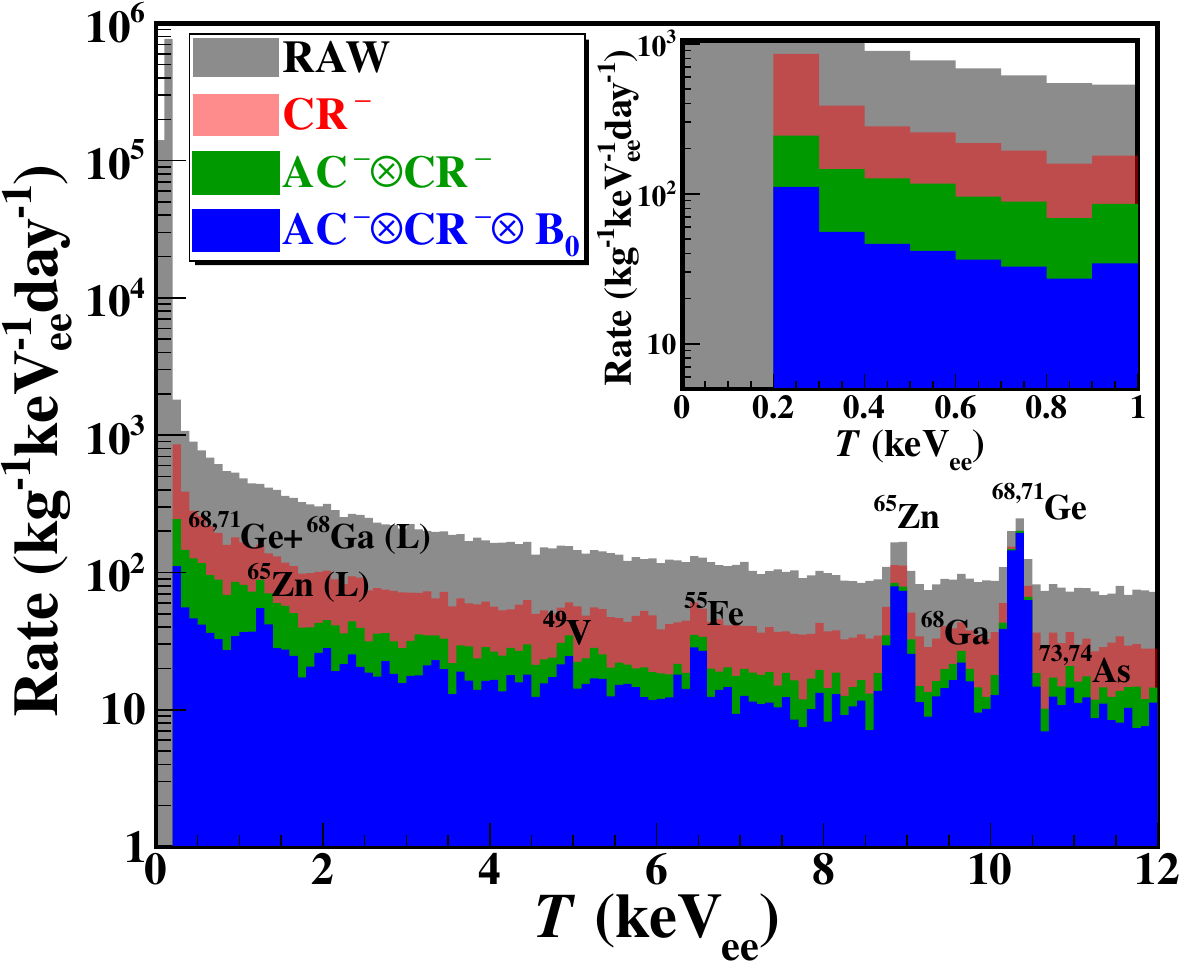}
  \end{center}
  \caption{Characteristic evolution of the
    measured spectra following the selection
    procedures described in
    Section~\ref{sect::analys}, using $\BigDet$
    as an illustration. Background
    events are suppressed to obtain the
    candidate samples. The near threshold
    behavior is expanded in the inset.}
    \label{fig::BkgLevel}
\end{figure}

Typical evolution of the measured spectra
following the selection procedures described
in Section~\ref{sect::analys} is depicted
in Figure~\ref{fig::BkgLevel}, which
demonstrates the effectiveness and energy 
dependence of each selection step in
rejecting background events. The RAW data
incorporate only the minimal baseline
selection. Representative spectra are
shown to illustrate the progressive
reduction of background through the successive
application of selection criteria. The
RAW data are followed by the $\CRV$,
$\VV$, and finally the $\nuAel$
candidate $\VVB$ spectra, which exhibit
a threshold of $\T0 {=} 200~ \eVee$. X-ray peaks from
cosmic-ray-induced isotopes  are also
identified.

\subsubsection{Consistency checks with different trigger samples}
\label{sect::trig::check}
As noted in Section~\ref{sect::analys}, together
with the RT and test pulser events, the different
background samples as tagged by various
combinations of CR$^{+}$, AC$^{+}$, and $S$ are used
for {\it in situ} efficiency evaluation and optimization of the
analysis software parameters. In
addition, their measurements are required to be
consistent with physics expectations so as to
provide independent cross-checks of the overall
performance of the detector hardware and data
analysis procedures. Two features are presented in the
following as illustration.


\begin{figure}[h!]
  \begin{center}
{\bf (a)}\\
\includegraphics[width=8.2cm]{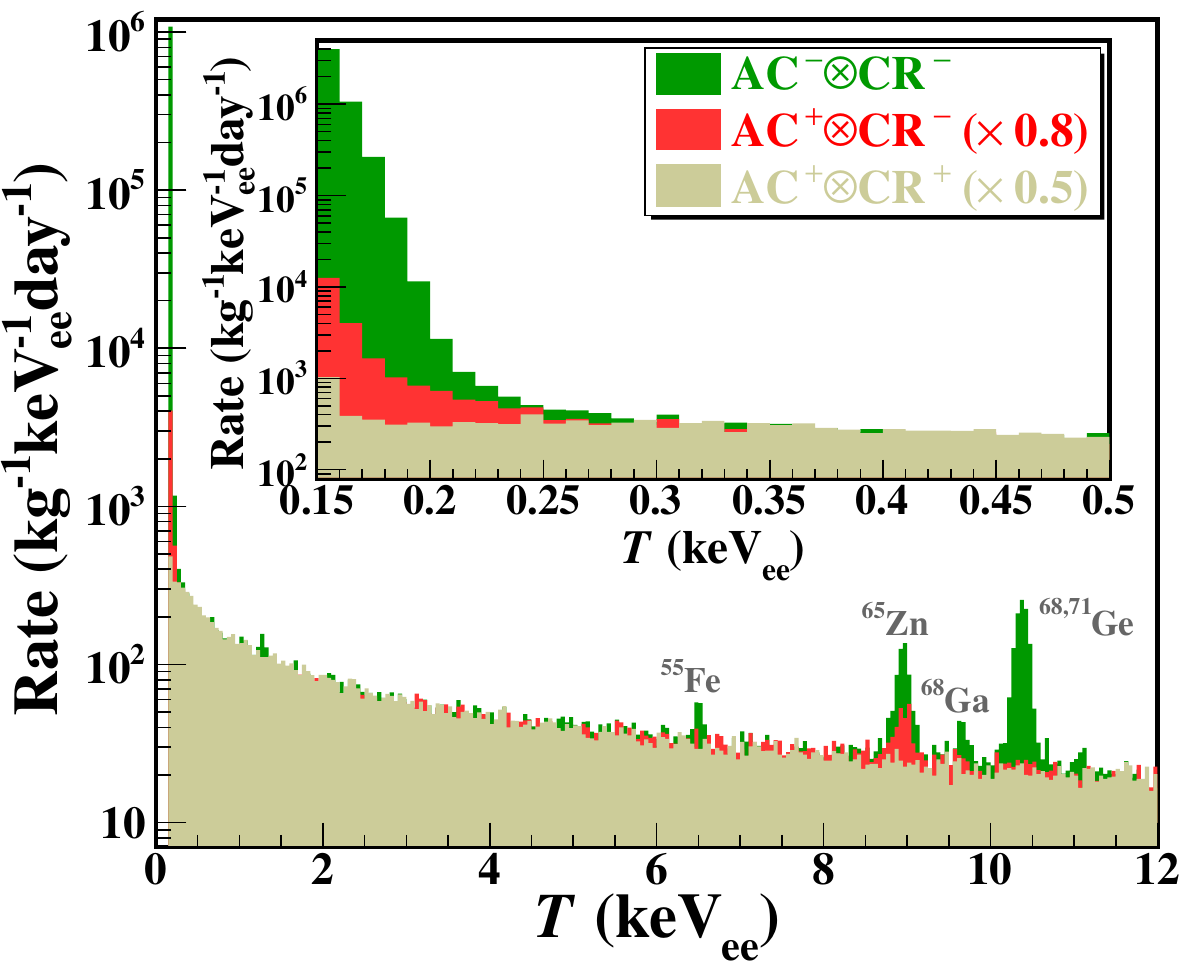}\\
{\bf (b)}\\
\includegraphics[width=8.2cm]{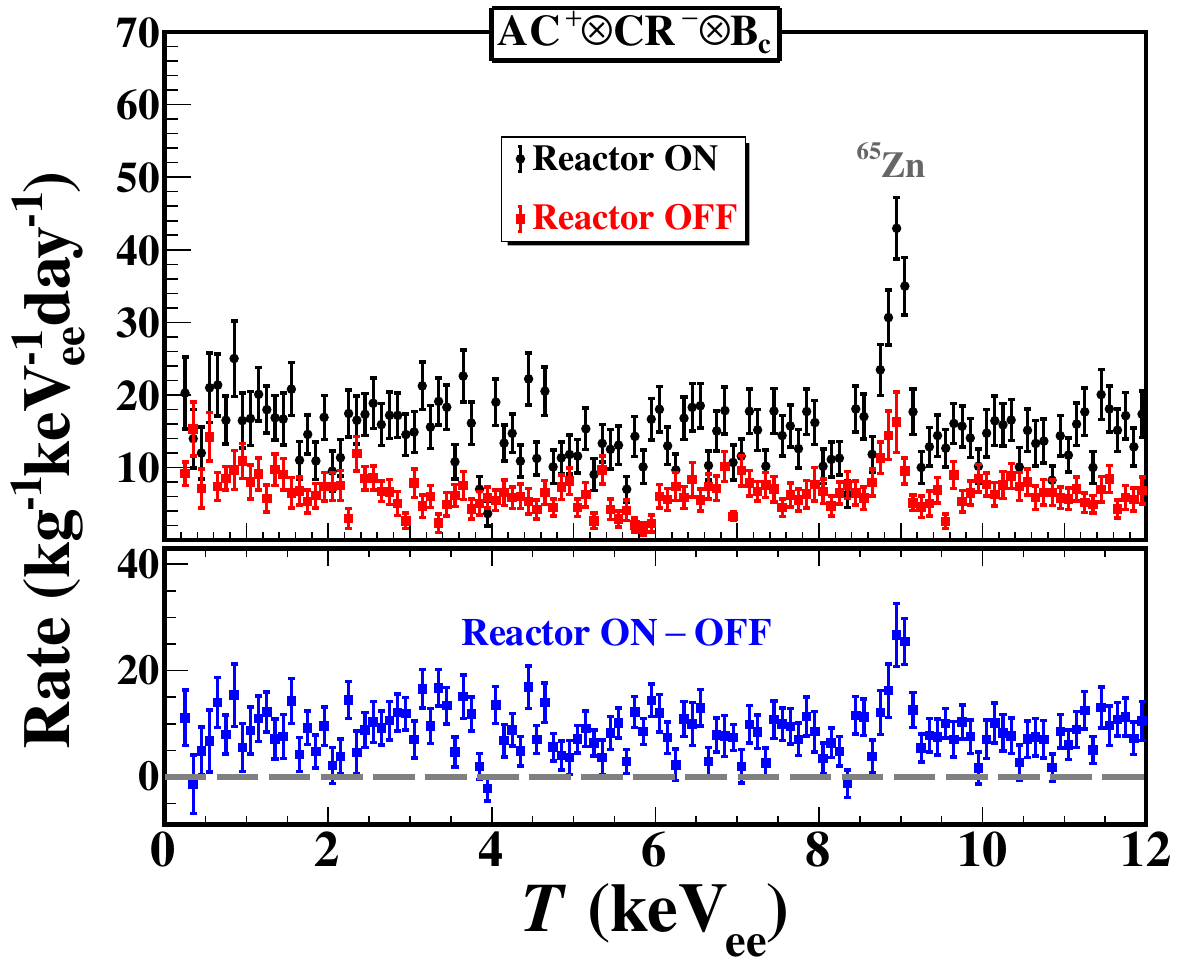} 
  \end{center}
  \caption{(a) Typical spectra for $\VV$, $\VT$, and
    $\TT$ events, using $\BigDet$ as an illustration,
    showing the reduction of NE as the
    coincidence requirements increase. (b) Typical
    Reactor ON, OFF, and ON$-$OFF spectra for
    $\VTBc$~\cite{Yang:2016crf}, using $\BigDet$
    as an illustration, which confirm
    predictions that the events are dominated by
    high energy ambient $\gamma$ rays undergoing Compton
    scattering in the detectors. The energy
    distributions are flat at low
    energies~\cite{LinST:2020}.}
  \label{fig::VTBcheck}
\end{figure}

Typical spectra for $\VV$, $\VT$, and $\TT$ samples
are depicted in Figure~\ref{fig::VTBcheck}(a).
Their high energy levels are normalized via
scaling factors for display purposes. It can be
seen that the spectral threshold, or
NE, decreases with increasing
coincidence requirements. It also follows that
 $\TT$ events provide valuable {\it in situ}
near- and sub-threshold physics samples for the
pulse shape analysis and efficiency
measurements of the $\VV$ $\nu$-candidate
events. The X-ray peaks are due to cosmogenic
activation of long-lived isotopes. They can
serve as {\it in situ} calibration measurements
and provide stability monitoring.

The $\VTBc$ samples have been processed under
identical procedures as the $\VV$ $\nu$-candidate
events $-$ with the exception of positive
signatures in the NaI(Tl) AC detector. Typical
spectra for Reactor ON, OFF, and ``residual''
ON$-$OFF are displayed in
Figure~\ref{fig::VTBcheck}(b). A flat spectral
shape at low energy is observed, consistent
with predictions~\cite{LinST:2020} that
these samples are dominated by Compton scattering
of ambient high energy $\gamma$ rays in the
Ge-detector. This feature confirms the validity
of the analysis procedures and parameter
choices, in particular the $B/S$
event selection and subsequent correction
from B$_{0}$ to B$_{\rm c}$. The expected
X-ray peak arises from $^{65}$Zn, which decays
with a half-life of 244~days and a branching
ratio of 50.08$\pm$0.06\% in coincidence
with the emission of high energy $\gamma$
rays. All other cosmogenically activated
isotopes in Ge decay via electron capture,
followed by the emission of single low
energy X-rays. Accordingly, only the $^{65}$Zn
peak would appear in the $\VTBc$ spectra.

\subsubsection{Candidate spectra}

\begin{figure}
  \begin{center}
  {\bf (a)}  \\
  \includegraphics[width=8.2cm]{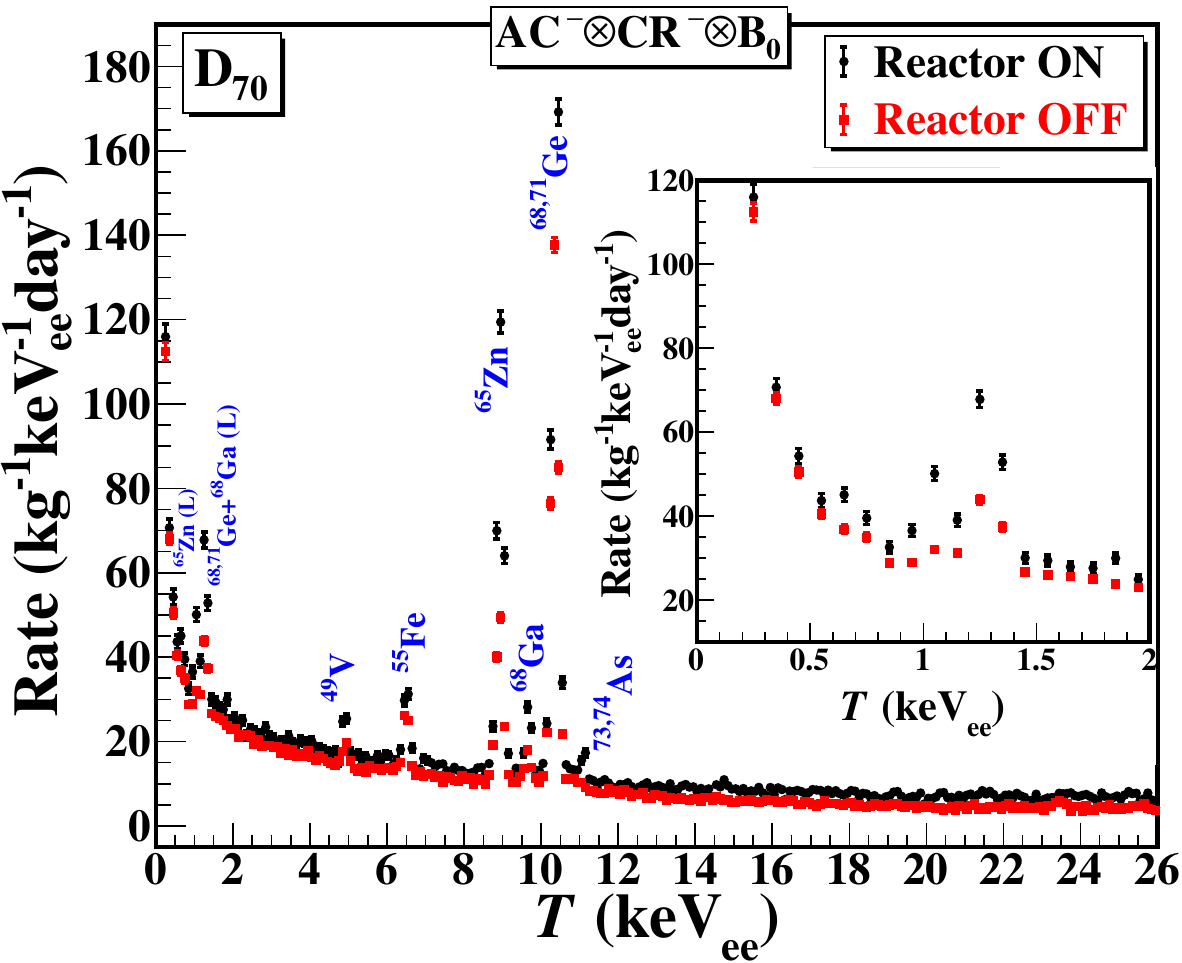} 
  {\bf (b)}  \\
  \includegraphics[width=8.2cm]{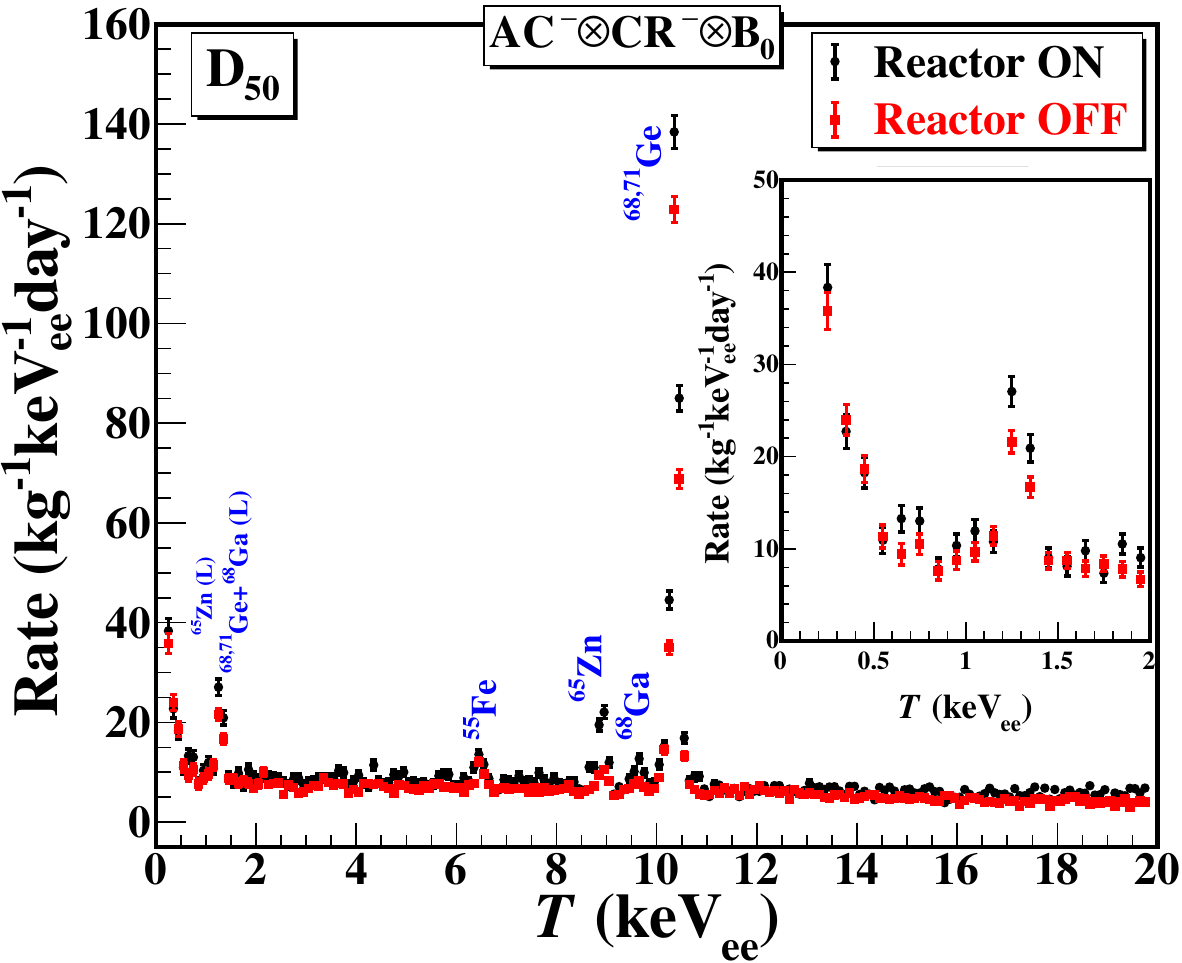}  
  \end{center}
  \caption{Measured Reactor ON and OFF spectra of
    $\VVB$ signal events from the full datasets of
    (a) $\BigDet$ and (b) $\SmallDet$ ECPCGe
    detectors, after applying all
    analysis procedures discussed in
    Section~\ref{sect::analys}. The insets show
    the low energy region below 2~$\keVee$, extending
    down to the analysis $\T0{=}$200~$\eVee$.
    The two detectors exhibit distinct high energy
    dynamic ranges, extending up to 26~$\keVee$ and
    20~$\keVee$ for $\BigDet$ and $\SmallDet$,
    respectively. The cosmogenic X-ray peaks are
    identified.}
  \label{fig::ON+OFF}
\end{figure}

The spectra of $\VVB$ candidate signal events 
from the full datasets after applying the
selection procedures described in Section~\ref{sect::analys} are
displayed in Figures~\ref{fig::ON+OFF}(a) and \ref{fig::ON+OFF}(b) 
for energies up to 26~keV$_{\rm ee}$ and 20~$\keVee$ for $\BigDet$ and
$\SmallDet$, respectively. The Reactor ON(OFF) exposures
are 242(559.3)~$\kgd$ and 162(254.4)~$\kgd$ for $\BigDet$
and $\SmallDet$, respectively. The analysis threshold is
200~$\eVee$.

The region below 12~$\keVee$ is populated by the
characteristic X-ray peaks originating from decays of
isotopes produced by cosmic-ray-induced interactions.
These are valuable for calibrating and characterizing
the energy response and resolution of the ECPCGe detector.


A note on analysis strategy should be made.
Reactor experiments are equipped with a powerful
selection cut that is background-model
independent $-$ taking residual spectra with
Reactor ON$-$OFF data. In this analysis, optimal
sensitivity is achieved using the $B/S$
selection shown in Figure~\ref{fig::cr+bs}(b),
and the residual spectra are derived in
$\VVB$. Software tools to correct B$_{0}$ to
B$_{\rm c}$ in order to account for $S$-events
leakage at low energy, do exist~\cite{TEXONO:2013bju,
  Yang:2016crf,Wang:2024phr}, and they are
essential in other analysis such as DM
searches. However, the residual spectra
in $\VVBc$ are not adopted in reactor
$\nuebar$ analysis because they incur larger
uncertainties in the low energy, near-threshold
bins and therefore reduce sensitivities of
the physics results.


\begin{figure}
  \begin{center}
    {\bf (a)}\\
    \includegraphics[width=8.2cm]{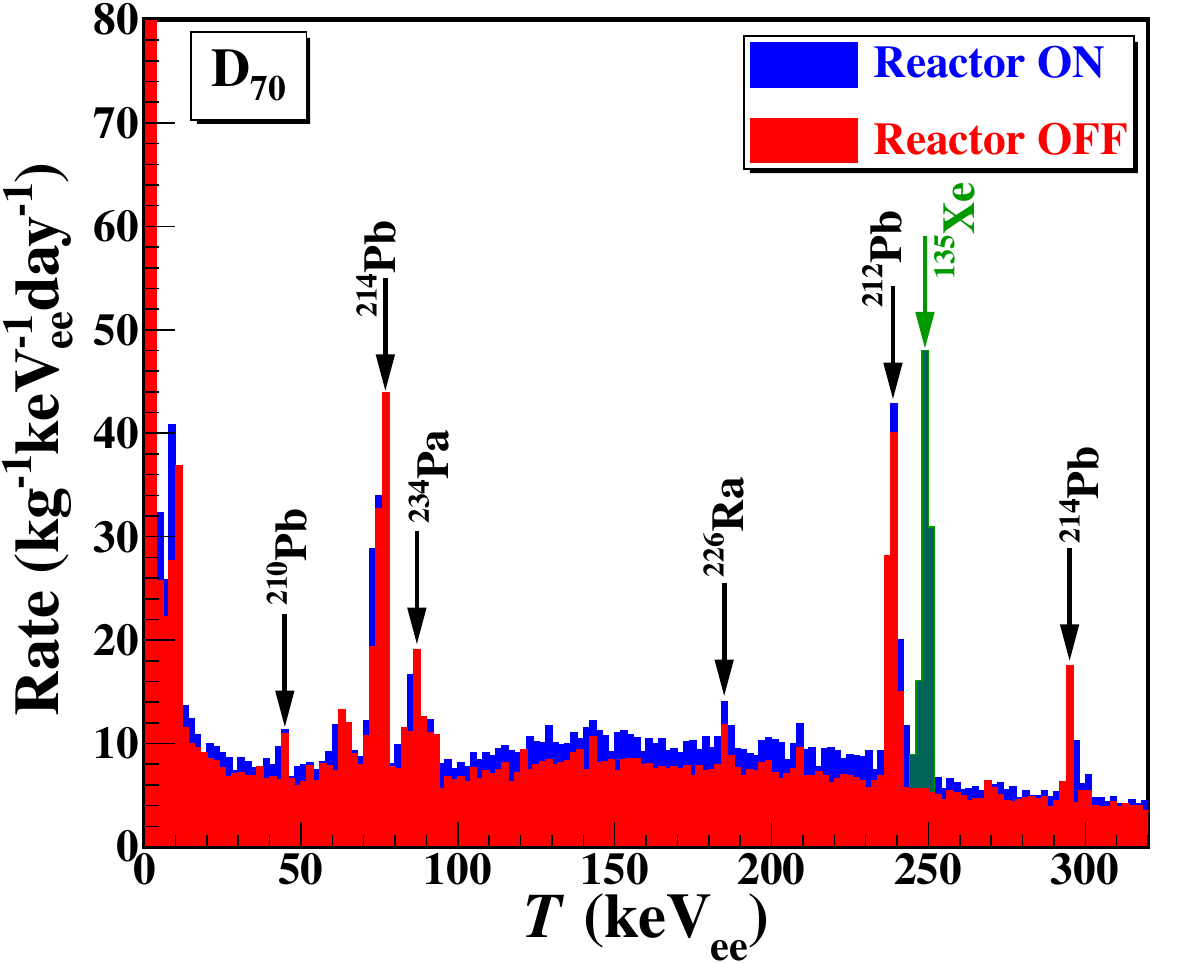} \\
    {\bf (b)}\\
    \includegraphics[width=8.2cm]{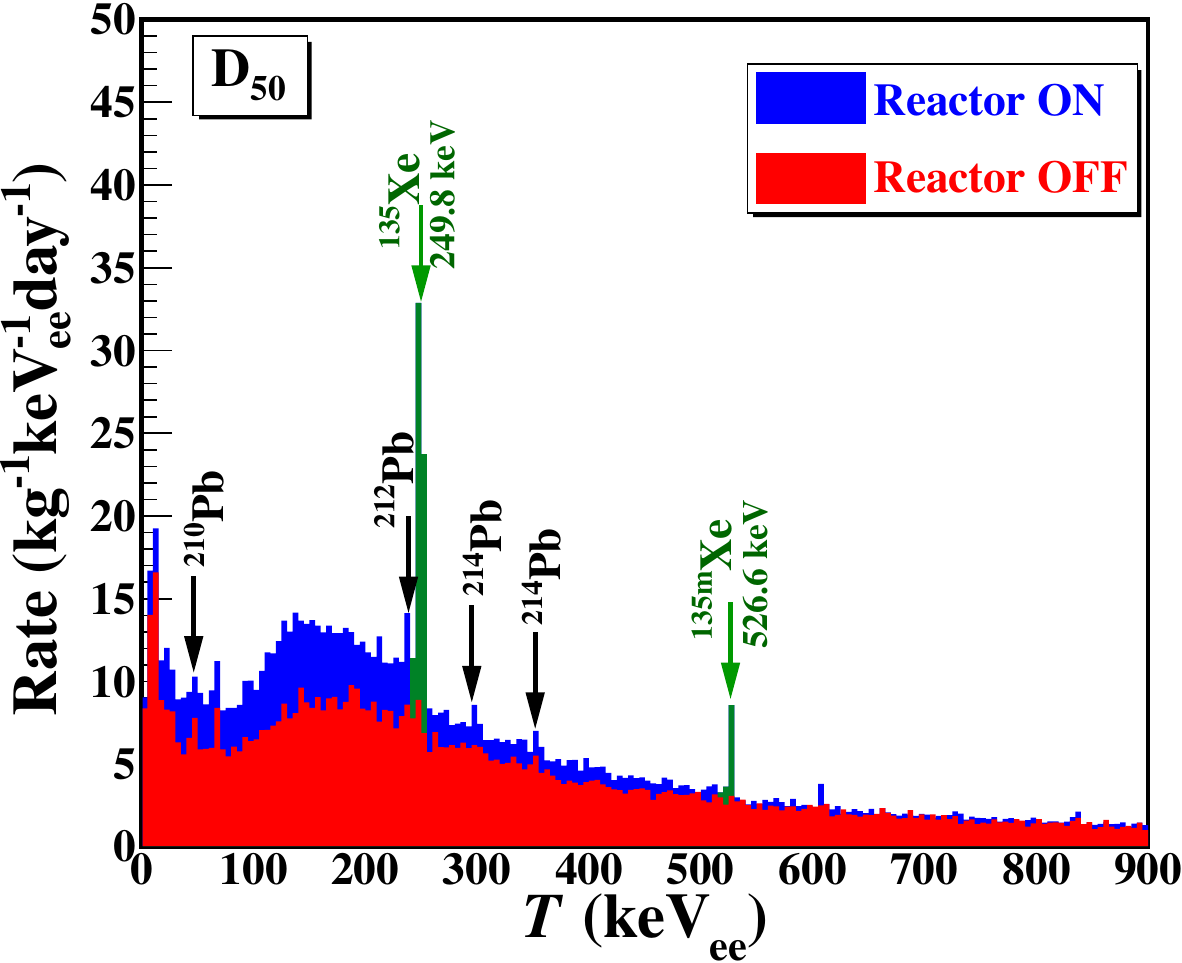} \\
    {\bf (c)}\\
    \includegraphics[width=8.2cm]{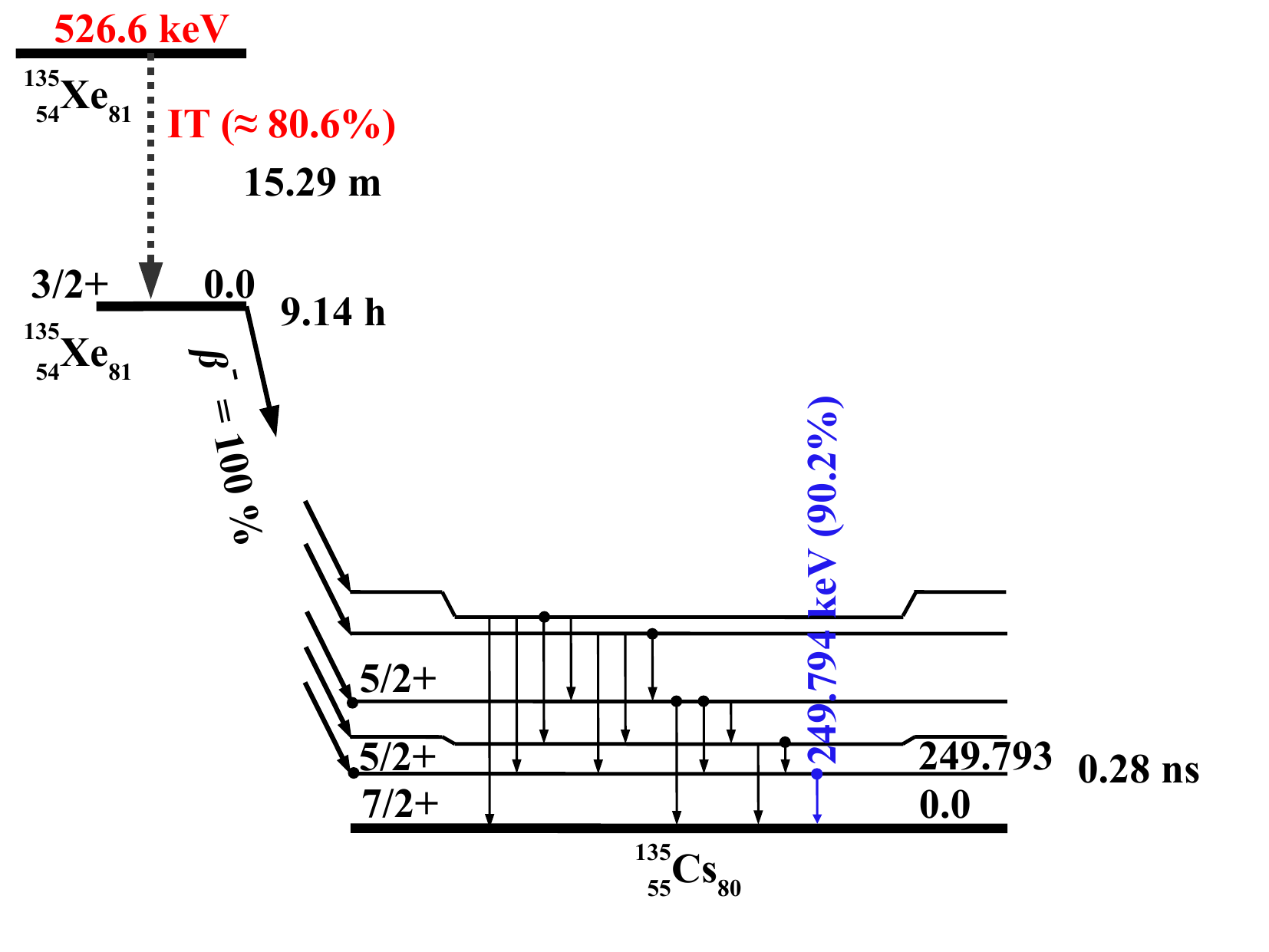}
  \end{center}
  \caption{Typical $\VV$ high energy spectra of the
    (a) $\BigDet$ and (b) $\SmallDet$ detectors
    during the Reactor ON and OFF periods. The 249.8~$\keVee$
    and 526.6~$\keVee$ $\gamma$-lines from $^{135}$Xe
    appear only in the Reactor ON data.
    (c) The decay scheme of $^{135}$Xe. It follows
    from the short half-lives that the background
    becomes negligible about one day after the reactor
    is switched OFF.}
  \label{fig::Xebkg}
\end{figure}

\subsection{Reactor ON $\xe135$ Background}
\label{sect::xe::bkg}

\begin{figure}
  \begin{center}
    {\bf (a)} \\
    \includegraphics[width=8.2cm]{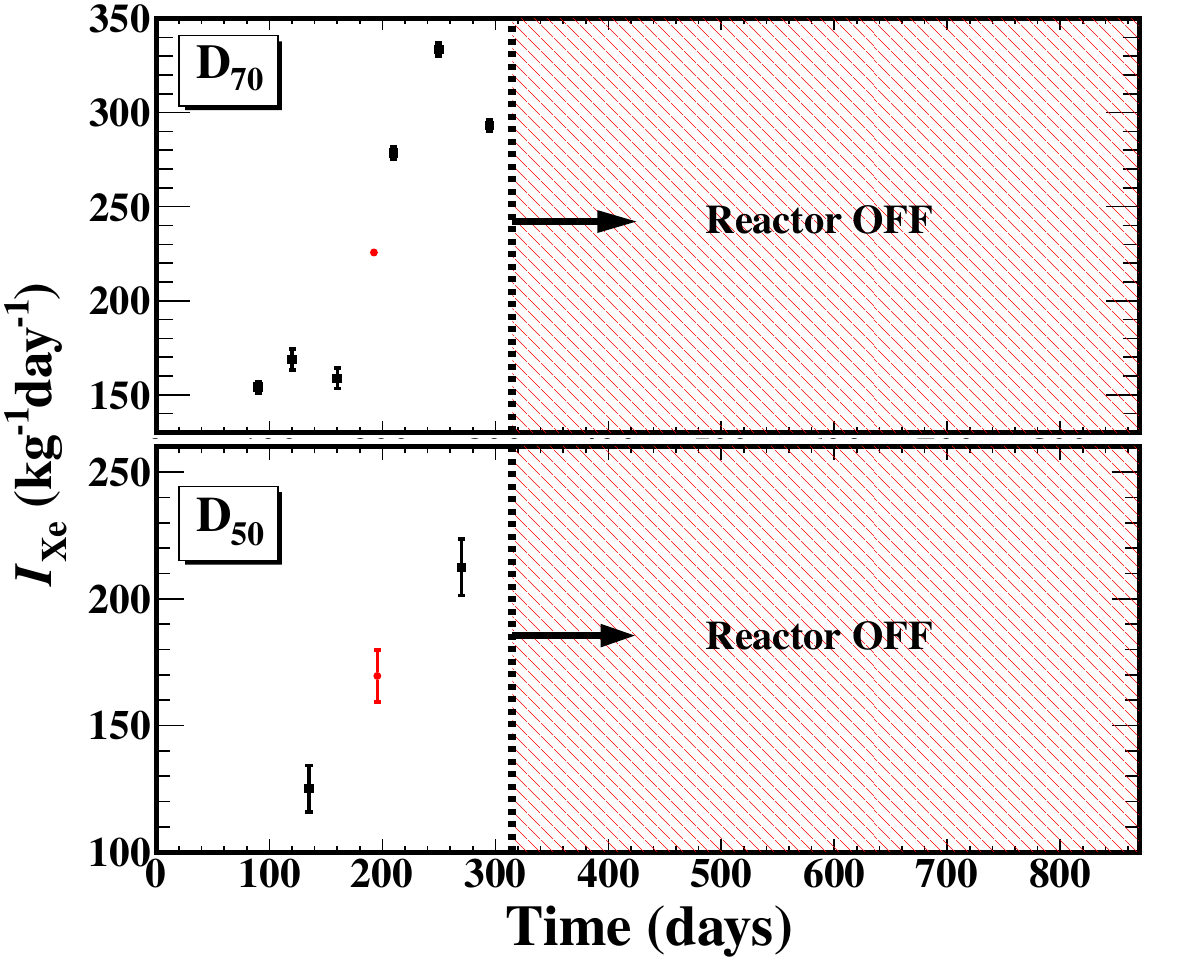} \\
    {\bf (b)} \\
    \includegraphics[width=8.2cm]{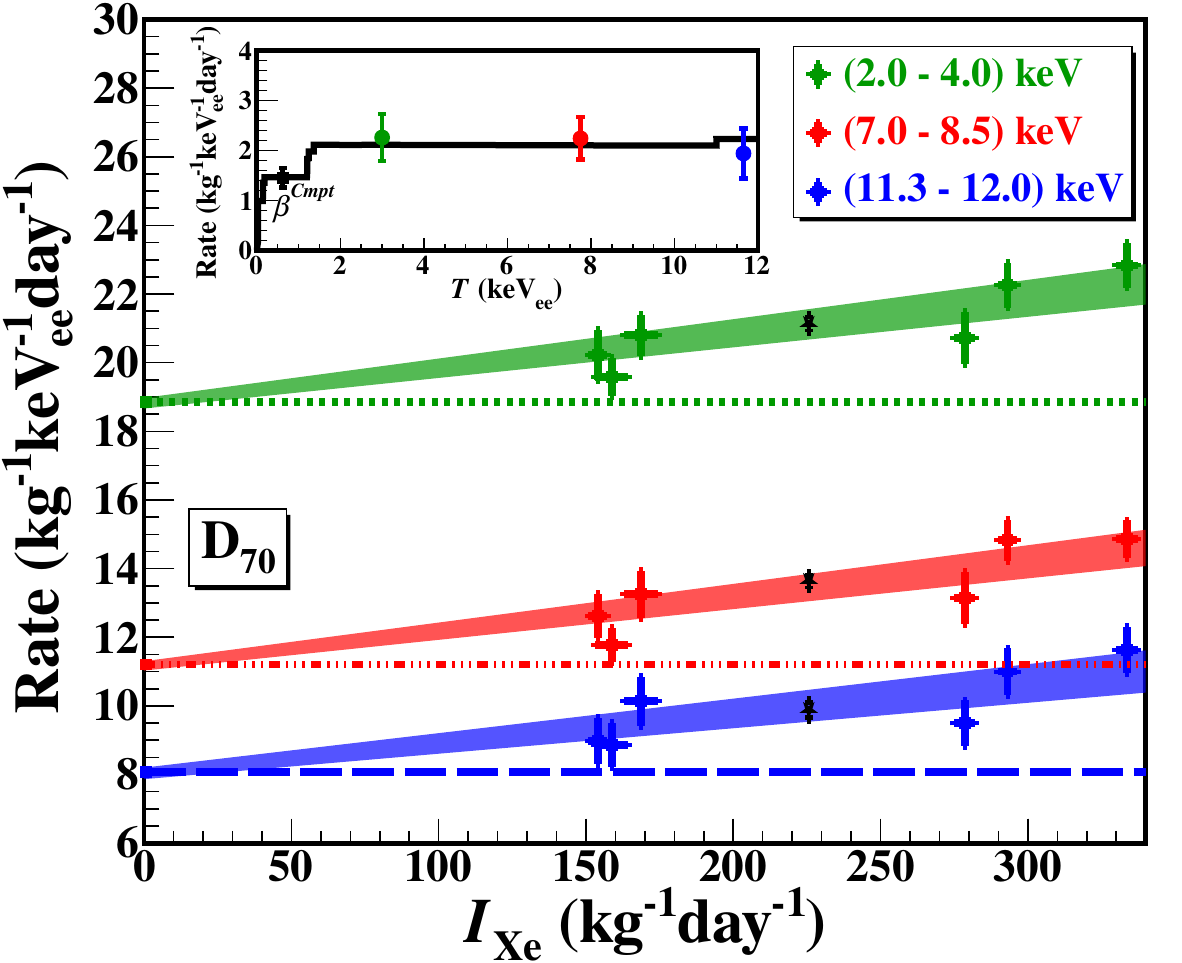} 
  \end{center}
  \caption{(a) Time evolution of the $I_{\rm Xe}$
    background, as indicated by the measured
    249.8~$\keVee$ $\gamma$-lines. (b) Using $\BigDet$
    as an illustration, the correlation between the
    constant background rates in three selected
    low energy bands, which are free of cosmogenic
    X-ray peaks, and $I_{\rm Xe}$. The observed
    linear relationship indicates that variations
    in the Reactor ON low energy background can
    be directly attributed to the high energy
    $\gamma$ activity from $\xe135$. The intercepts
    at $I_{\rm Xe}{=}$0 match the baseline Reactor OFF
    rates. The black data points indicate the
    measured values for the combined data, and
    the shaded region illustrates the corresponding
    $\pm$1$\sigma$ uncertainty bands. The inset
    displays the predicted Compton
    background~\cite{LinST:2020,Hu::CDEX2023}, determined by
    the measured excess rates over the Reactor OFF
    rates in the three low energy bands.}
  \label{fig::xe-d1+2}
\end{figure}

An anomalous background during Reactor ON periods
is observed in both $\BigDet$ and $\SmallDet$,
characterized by the 249.8~keV $\gamma$-line and
attributed to $\xe135$ contamination, as
shown in Figures~\ref{fig::Xebkg}(a) and \ref{fig::Xebkg}(b), respectively.
The line has been observed in our earlier
work~\cite{TEXONO:2007xds} at a lower intensity.
The isotope $\xe135$ is a fission product and
a strong neutron poison due to its large neutron
absorption cross section~\cite{SANTRY::1973::CS,
  DOE::HDBK::v2:1993}. It diffuses as gaseous
xenon into the inner space of the shielding
structure. The decay scheme is depicted in
Figure~\ref{fig::Xebkg}(c). In particular, the
strongest $\gamma$-line at 249.8~keV has a
branching ratio of approximately 90.2\% per
decay. The dynamic range of $\SmallDet$
extends to 900~keV. The much weaker high
energy $\gamma$-lines are also observed, as
shown in Figure~\ref{fig::Xebkg}(b).

The constraints on access to KSNL during the
pandemic periods imply that the background
cannot be suppressed {\it in situ}
and therefore requires to be accounted for
in offline analysis. The short half-life of
9.14~h allows its effects to be studied in
the Reactor ON$-$OFF measurements. The
observed intensities of the 249.8~keV $\gamma$-line
($I_{\rm Xe}$) depend on the time-varying reactor
operating conditions and shielding configurations, and
are displayed in Figure~\ref{fig::xe-d1+2}(a) for both
detectors. The potential background in the $\nuAel$
measurements is the $\xe135$-induced Compton continuum
at sub-keV energies. It is proportional to $I_{\rm Xe}$,
which is measured {\it in situ}, and therefore can be
predicted accurately in the analysis~\cite{TEXONO-PRL2025}.
As illustrated in Figure~\ref{fig::xe-d1+2}(b) with
$\BigDet$ data, the correlations are evaluated for the
2-4~keV, 7-8.5~keV, and 11.3-12~keV energy bands,
which are free from contributions of cosmogenic X-ray
peaks. The averaged rates in the three bands, indicated
by black markers, are consistently above their
respective Reactor OFF baseline levels defined at
$I_{\rm Xe}{=}$0. The predicted Compton
background~\cite{LinST:2020,Hu::CDEX2023}, with detector resolution
and atomic level structures incorporated and constrained
by the measured rates in the three bands, is displayed
in the inset of Figure~\ref{fig::xe-d1+2}(b).

\subsection{Reactor ON$-$OFF Spectra}
\label{sect::on-min-off}

\begin{figure} 
  \begin{center}
  {\bf (a)}  \\
  \includegraphics[width=8.2cm]{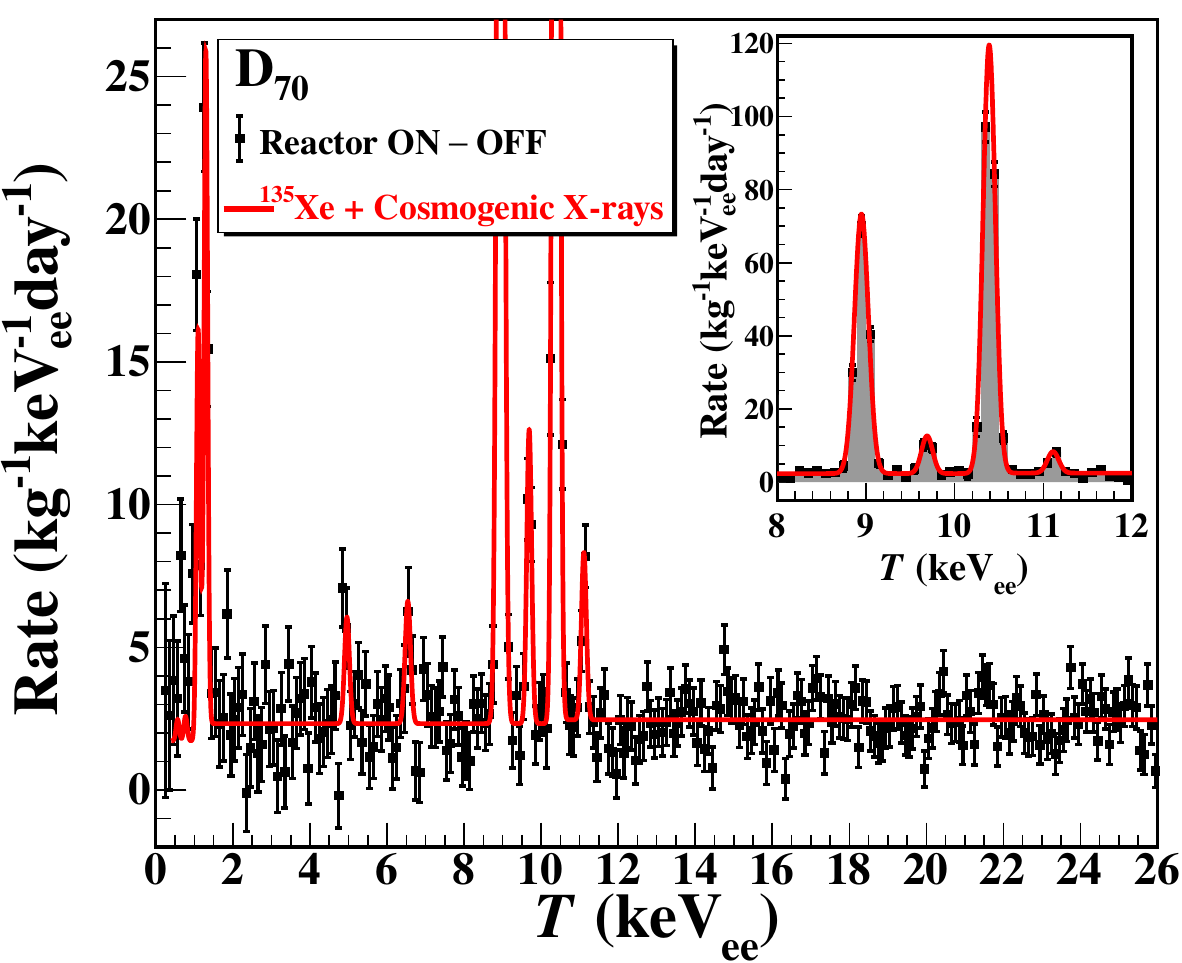}\\
  {\bf (b)}  \\
  \includegraphics[width=8.2cm]{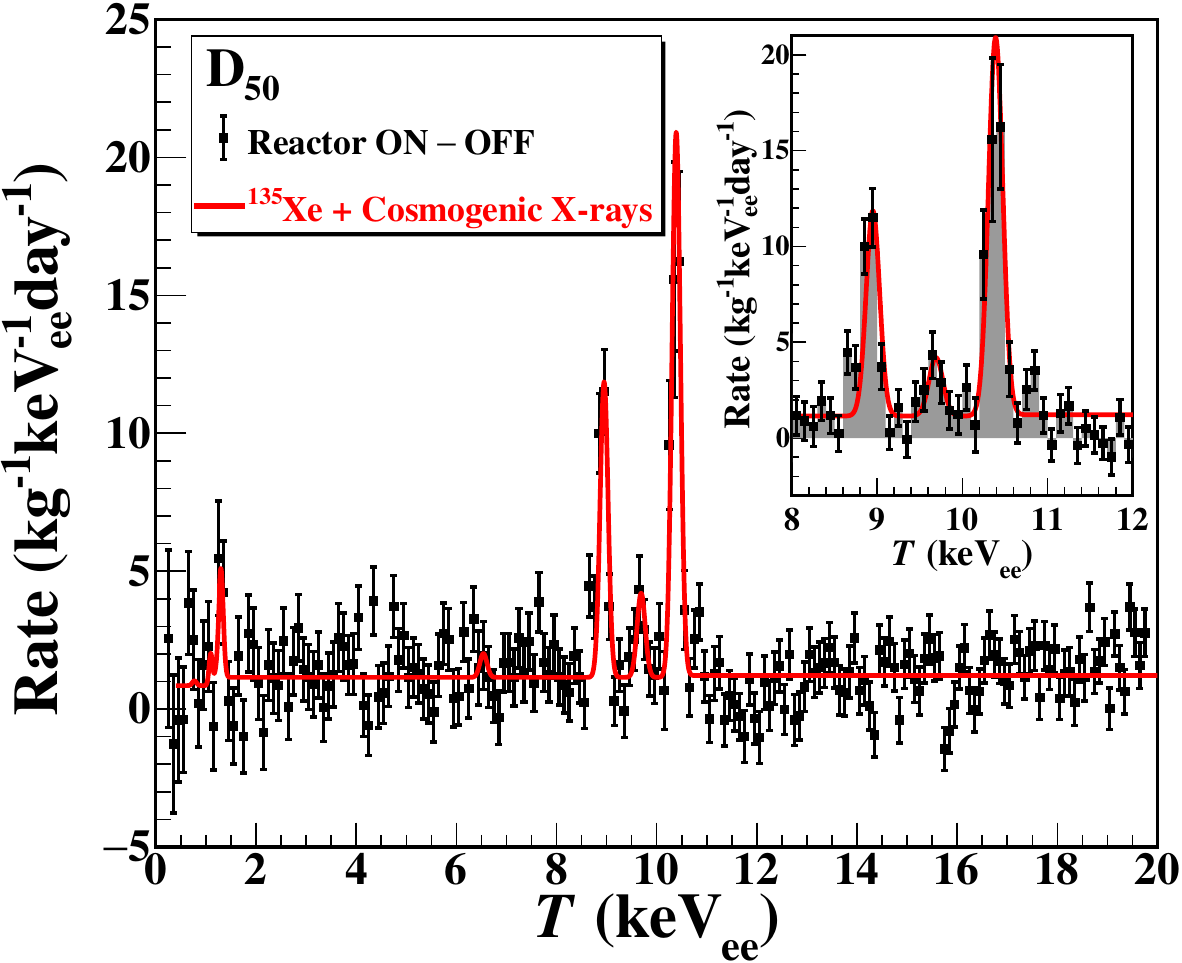}  
  \end{center}
  \caption{The Reactor ON$-$OFF residual
    spectra of the full datasets for
    (a) $\BigDet$ and (b) $\SmallDet$,
    showing the finite
    but minor constant background due to
    $\xe135$, as well as the X-ray peaks
    due to cosmogenic activities.
    The best-fits including both background
    components are shown as red. For
    clarity in display, the intense
    cosmogenic $^{68,71}$Ge 10.37~keV X-ray
    peak is truncated in (a).}
  \label{fig::residual}
\end{figure}

The Reactor ON$-$OFF residual spectra 
for the full datasets of $\BigDet$ and $\SmallDet$
are shown in
Figures~\ref{fig::residual}(a) and \ref{fig::residual}(b), respectively. 
The flat continuum is due to the
reactor-induced $\xe135$ background,
as discussed in Section~\ref{sect::xe::bkg}. 
Reactor ON data taking preceded Reactor
OFF for these datasets, therefore,
X-rays emitted by cosmogenic
isotopes with lifetimes in the range of
100-1000~days manifest as peaks in the
ON$-$OFF spectra. Both background are
taken into account and the best-fit
spectra are denoted in red.

In the relevant energy range below 400~$\eVee$,
the best-fit values of the $\xe135$
Compton background (denoted by $\beta^{\rm Cmpt}$)
are 1.72$\pm$0.02 (0.85$\pm$0.04)~events~$\pkkd$ for
$\BigDet (\SmallDet$). This corresponds to
less than 0.92(1.4)\% of the total Reactor
ON background rate for $\BigDet (\SmallDet$). The $\xe135$
contribution, though minor, is 
accurately quantified in the Reactor
ON$-$OFF data analysis from which the
physics results are derived.


\subsection{Physics Results}
\label{sect::phys::res}

The physics results are derived from the
combined data with the two Reactor ON$-$OFF
residual spectra from both detectors,
$\BigDet$ and $\SmallDet$. The total
exposure is 404(813.7)~$\kgd$ for Reactor
ON(OFF). A minimum $\chi^{2}$ analysis is
applied to the combined spectra in the
signal window of 200$-$400~$\eVee$ at
bin-size of 10~$\eVee$:
\begin{equation}
  \begin{aligned}
    \chi^{2} ( \rho , \beta, \munu ; k)  =  \\
    \sum_{i}\left[ \frac{N_{i}-(\rho ~ \nu^{\rm SM}_{i}(k)+\nu^{\rm BSM}_{i}(\munu))- \beta} {\Delta_{i}}   \right]^{2} \\
    + \left[ \frac{\beta  -\beta^{\rm Cmpt}}{\Delta^{\rm Cmpt}} \right]^{2}~,
    \label{eqn:chi2}
  \end{aligned}
\end{equation}
where $N_{i}$($\Delta_{i}$) are the
counts(uncertainties) of the $i^{th}$
data bin, $\nu^{\rm SM}_{i}$($k$) and
$\nu^{\rm BSM}_{i}$($\munu$) are the expected
counts from SM-$\nuAel$ and BSM-$\munu$
interactions, respectively, $\rho$
denotes the excess over the SM-$\nuAel$
processes, and $\beta$ represents the
$\xe135$ level of the combined datasets.

\begin{figure}
  \begin{center}
  {\bf (a)}\\
  \includegraphics[width=8.2cm]{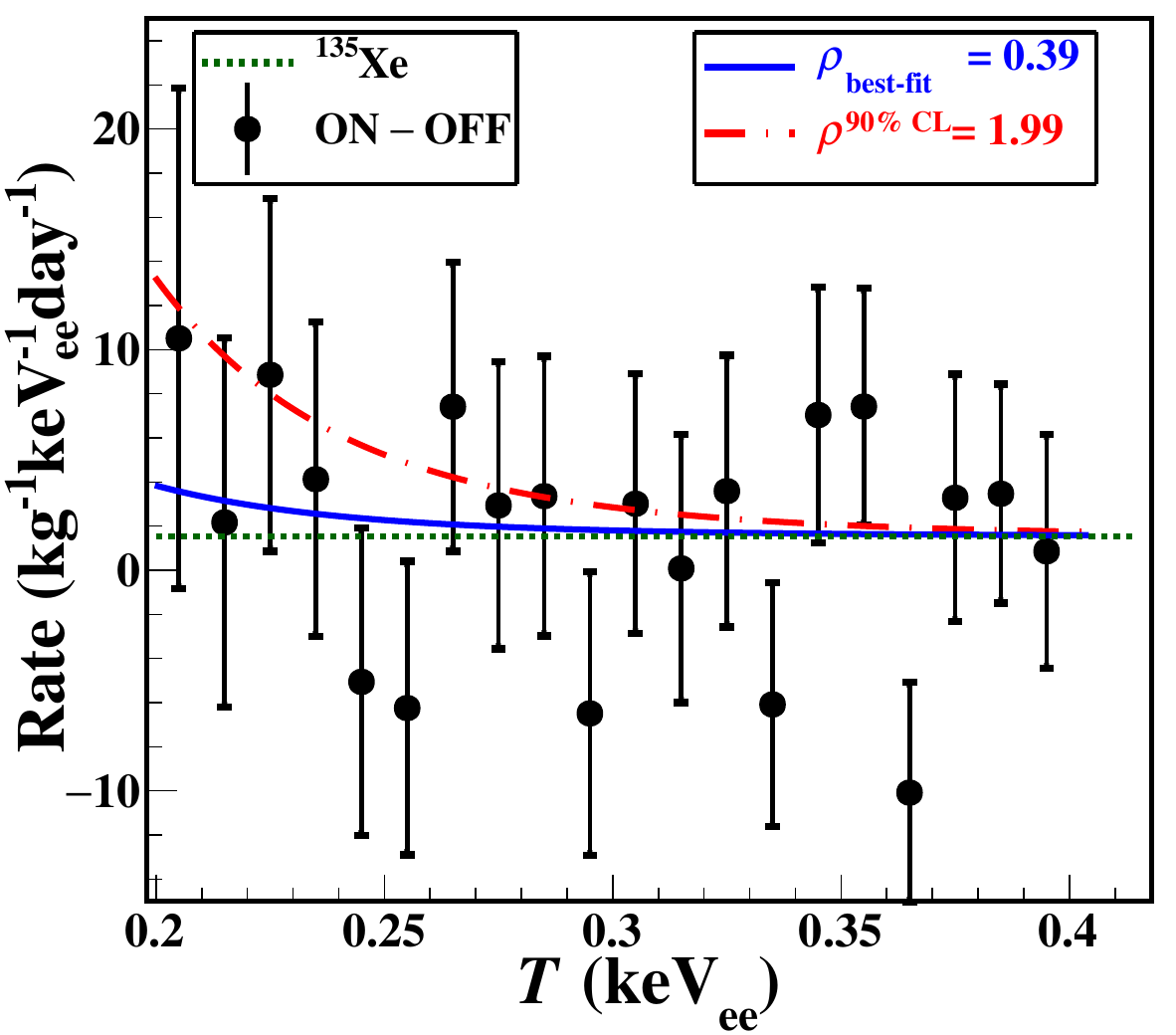} \\
  {\bf (b)}  \\
  \includegraphics[width=8.2cm]{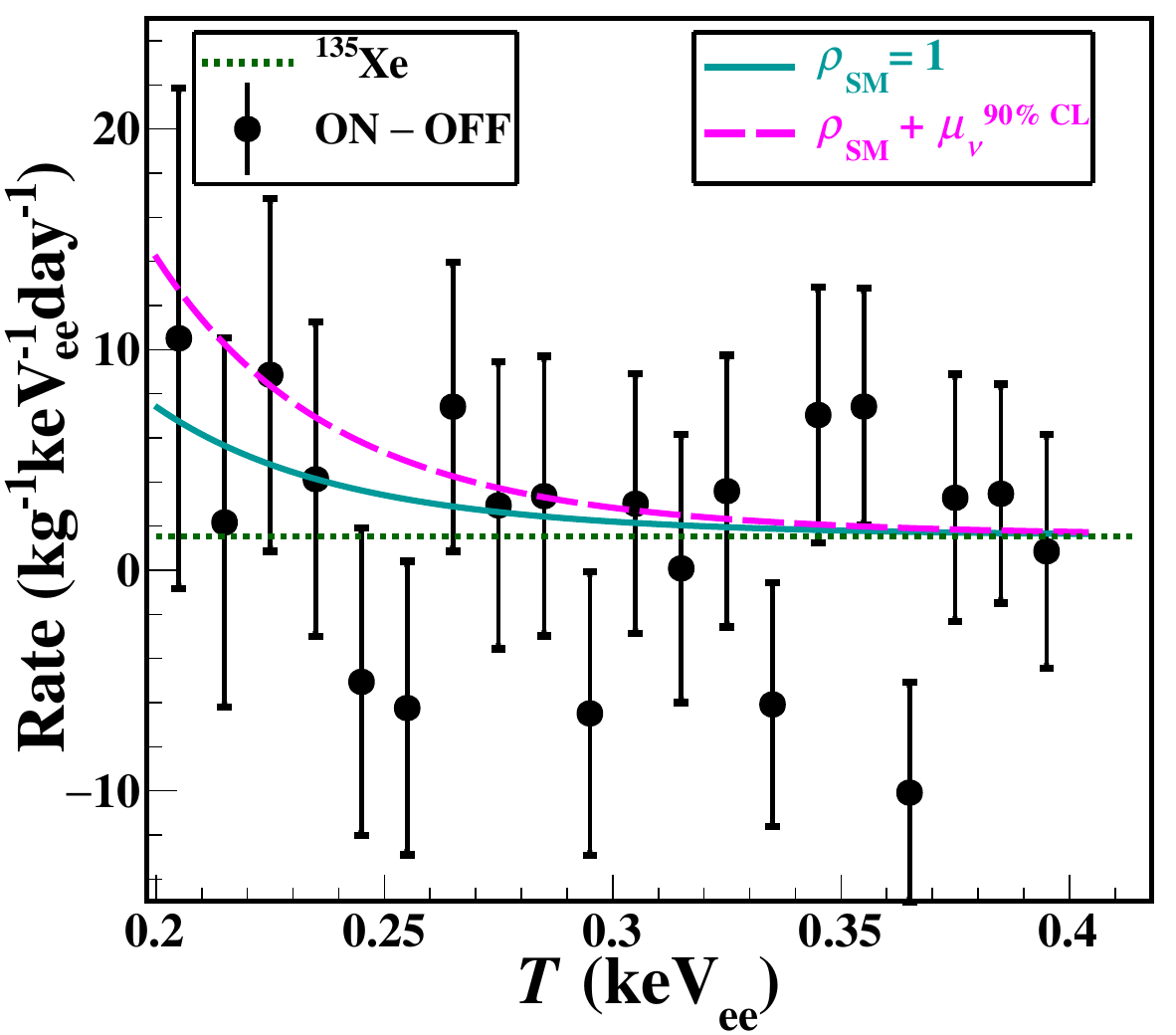} 
  \end{center}
  \caption{Reactor ON$-$OFF residual spectrum for
    the combined $\BigDet$ and $\SmallDet$ full datasets
    for $\VVB$ events, showing (a) the $\nuAel$
    channel for the best-fit and the 90\% CL limit
    on $\rho$, and (b) the $\munu$ channel with
    $\nuAel$ fixed at $\rho${=}1, along with the
    90\% CL limit on $\munu$. The $\xe135$
    background is represented by the green dotted line.}
  \label{fig::nuAfit}
\end{figure}

\subsubsection{Standard Model $\nuAel$ } 
At $k$=0.162, as adopted in most
analyses, the best-fit values and
their associated uncertainties for $\munu$=0 are
\begin{eqnarray}
  \beta~{=}~1.55 \pm 0.02 ({\rm stat.}),~{\rm and} \nonumber \\[1mm]
  \rho~{=}~0.39~^{+0.97}_{-0.39}~({\rm stat.})~^{+0.11}_{-0.07}~({\rm sys.}),
\end{eqnarray}
from which
\begin{equation}
  \rho~{<}~2.0,
\end{equation}
at 90\% CL is derived with the unified
approach~\cite{Feldman_PRD98,JNeyman:1935}.
Statistical uncertainties dominate the results at
the current level of sensitivity. Depicted in
Figure~\ref{fig::nuAfit}(a) is
the residual spectrum of the full datasets 
with the best-fit and the 90\% CL spectra at $k${=}0.162.
The increased error bar at the
$\T0{=}$200~$\eVee$ is due to the drop in
signal efficiency shown in
Figure~\ref{fig::sigeff}. The measured
$\beta$ is low relative to the spectral
uncertainties (which is 11.3~$\pkkd$ at
200~$\eVee$), indicating that the effects due
to Reactor-ON-induced $\xe135$ background 
are minor at the present level of
sensitivity.

\begin{figure}[h!]
  \begin{center}
  \includegraphics[height=7.2cm,width=8.2cm]{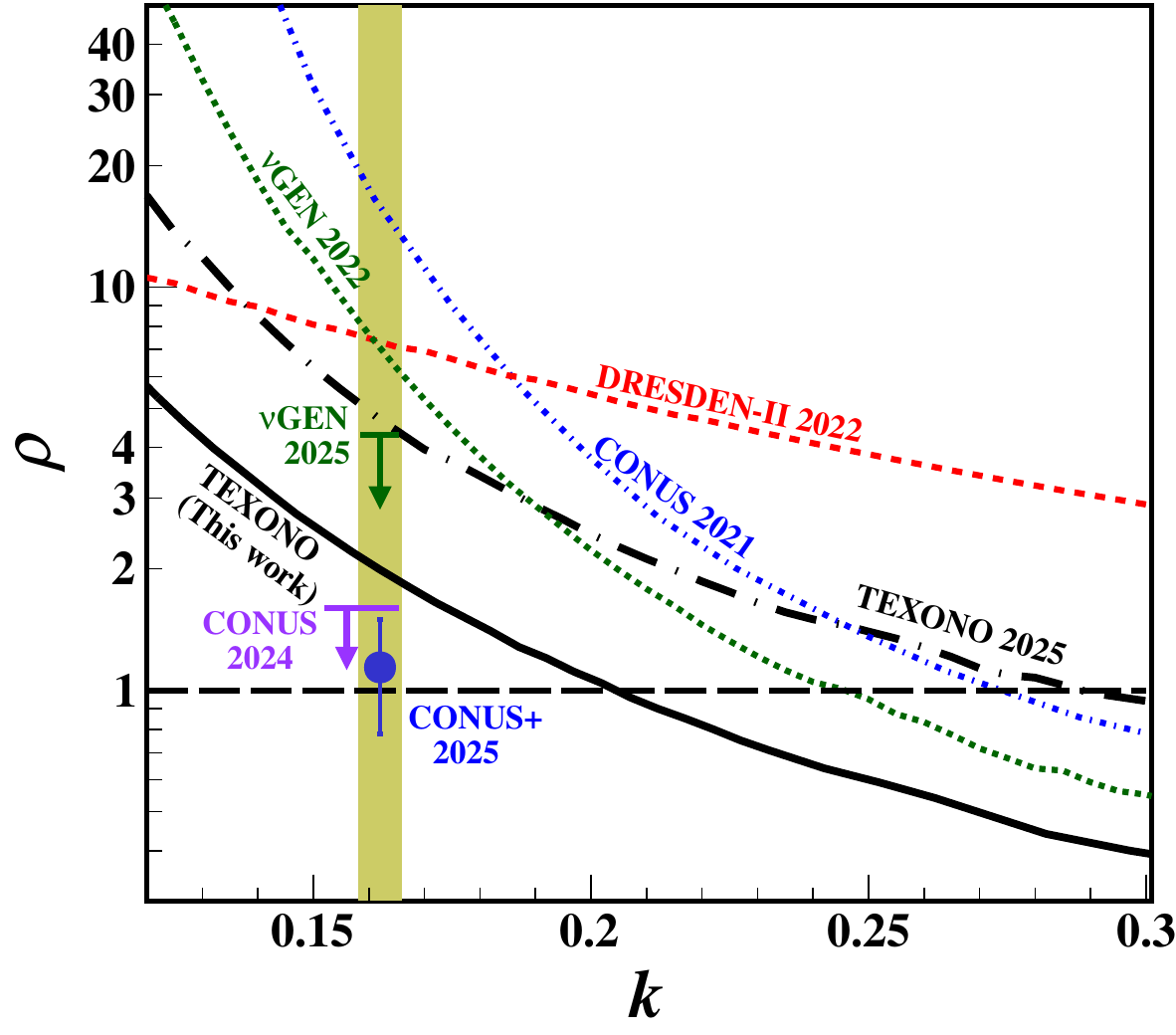} 
  \end{center}
  \caption{Exclusion plot in the ($\rho$, $k$)
    parameter space from the combined
    $\BigDet$ and $\SmallDet$ datasets.
    Superimposed are constraints from previous
    TEXONO results~\cite{TEXONO-PRL2025} and
    from other $\nuAel$ projects using
    Ge-detectors $-$ DRESDEN~\cite{DRESDEN-1},
    $\nu$GeN~\cite{nuGeN-2,nuGEN::CPC2025},
    CONUS~\cite{Bonet:2020awv,CONUS-2}, and
    CONUS+~\cite{CONUS+2025-1}.}
  \label{fig::explot}
\end{figure}

Figure~\ref{fig::explot} displays the
exclusion plot of $\rho$ versus $k$ at
90\% CL. Results from the other reactor
PCGe experiments are superimposed,
either as published
results~\cite{nuGEN::CPC2025,CONUS-2,
  CONUS+2025-1} or as derived with the
same analysis
procedures~\cite{DRESDEN-1,Bonet:2020awv,nuGeN-2}.
At the SM value of $\rho${=}1, an
upper limit of
\begin{equation}
  k~{<}~0.205,
\end{equation}
at 90\% CL is derived. This constraint is presented
in Figure~\ref{fig::qf} in the QF versus $\TNR$
space. It is evident that some of the outlying data points of 
Refs.~\cite{PhysRevC.4.125}\&\cite{JICollar_QF_PRD21}
are rejected. For completeness, the
latest result from the CONNIE $\nuAel$ project
using silicon skipper CCDs corresponds to a limit of
$\rho {<}$76 at 95\% CL~\cite{CONNIE::2024::arXiv}.

\subsubsection{Magnetic Moments $\munu$ }
As depicted in Figure~\ref{fig::nuAfit}(b),
at the SM configuration of ($\rho${=}1; $k${=}0.162),
no statistically significant excess associated
with a nonzero $\munu$ is observed. Consequently,
an upper limit on the $\munu$ is derived from
Eq.~(\ref{eq:A8}) as
\begin{equation}
  \munu~ {<}~5.9\times10^{-10} \muB,
\end{equation}
at 90\% CL. This result is comparable to the
recent constraints~\cite{Corona:PRD2025nuNCh}
derived from reactor $\nuAel$ data collected
by the CONUS+~\cite{CONUS+2025-1} and
TEXONO~\cite{TEXONO-PRL2025} experiments, which
reported limits of $\munu {<} 5.6\times10^{-10}$,
and ${<} 11\times10^{-10}\muB$, at
90\% CL, respectively.


\section{Summary and Prospects}
\label{sum:pro}

We report the final results on $\nuAel$ from the TEXONO
experiment at KSNL, using  the complete analyzable
datasets with ECPCGe detectors at a $\T0{=}$200~$\eVee$.
Experimental details are presented, and improved limits
are derived relative to our earlier work with a partial
dataset~\cite{TEXONO-PRL2025}.

The KS Nuclear Power Station was decommissioned in
March 2023, following a national policy of phasing out
nuclear power. Through special arrangements, KSNL
remains operational. A good-quality,
well-characterized shielding structure is in place for
background studies and detector R\&D programs.
Feasibility reviews were conducted in late 2025
toward a possible restart of nuclear power generation
at KS, raising hopes that the neutrino physics
studies at KSNL may resume.

The TEXONO research program at KSNL will be upgraded
and continued with the RECODE project at the Sanmen
Reactor Neutrino Laboratory~\cite{Yang:20249T,Shen:2026kzy,RECODE::SChina::2025}
in Zhejiang, China, which is currently under
construction and commissioning. In addition, the
teams are studying next-generation ECPCGe detectors with
reduced thresholds, as well as advanced software
analysis techniques for near-and sub-threshold
pulse shape discrimination.

\section{Acknowledgment}
The data presented in this work were taken at KSNL
during the difficult periods of the COVID-19 pandemic.
The authors are indebted to all staff of our
institutes and of the Kuo-Sheng Nuclear Power Station
for their support in making this work possible. Funding of
this work was provided by the Investigator Award AS-IA-106-M02 
and Thematic Project AS-TP-112-M01 from Academia
Sinica, Taiwan, and by contracts 106-2923-M-001-006-MY5, 
107-2119-M-001-028-MY3, 110-2112-M-001-029-MY3, and 
113-2112-M-001-053-MY3 from the National Science and
Technology Council, Taiwan, and by the Taiwan International
Graduate Program in Physics (TIGP-X).

\bibliography{nulong-ref}

\end{document}